\documentclass{article}

\usepackage{arxiv}

\usepackage[utf8]{inputenc} 
\usepackage[T1]{fontenc}    
\usepackage{hyperref}       
\usepackage{url}            
\usepackage{booktabs}       
\usepackage{amsfonts}       
\usepackage{nicefrac}       
\usepackage{microtype}      
\usepackage{lipsum}		
\usepackage{graphicx}

\usepackage[square,sort,comma,numbers]{natbib}

\usepackage{doi}
\usepackage{tabularx}  
\usepackage{xltabular}
\usepackage{afterpage}
\usepackage{siunitx}
\usepackage[export]{adjustbox}

\usepackage{todonotes}
\usepackage{xcolor}
\usepackage[printonlyused,nohyperlinks]{acronym} 

\usepackage{orcidlink}

\title{Self-Perception Versus Objective Driving Behavior: Subject Study of Lateral Vehicle Guidance}

\author{%
	Johann Haselberger\,\orcidlink{0000-0002-2458-3461}\textsuperscript{*,$\dagger$,$\ddagger$}\, 
	Bernhard Schick\,\orcidlink{0000-0001-5567-3913}\textsuperscript{*}\, 
	Steffen Müller\textsuperscript{$\dagger$}\\
	\textsuperscript{*}Institute for Driver Assistance and Connected Mobility, University of Applied Science Kempten\\
	\textsuperscript{$\dagger$}Department of Automotive Engineering, Technical University Berlin\\
	\textsuperscript{$\ddagger$}Corresponding author. Tel.: +4917692612417\\
	\texttt{\{johann.haselberger, bernhard.schick\}@hs-kempten.de, steffen.mueller@tu-berlin.de}   
}

\renewcommand{\shorttitle}{Self-Perception Versus Objective Driving Behavior}

\hypersetup{
pdftitle={Self-Perception Versus Objective Driving Behavior},
pdfsubject={q-bio.NC, q-bio.QM},
pdfauthor={Johann~Haselberger, Bernhard~Schick, Steffen~Müller},
pdfkeywords={Driving Style, Driving Behavior, Subject Study, Human Factors, Self-Reports, MDSI, Instrumented Vehicle, German Drivers},
}

\usepackage{amssymb}
\usepackage{pifont}
\newcommand{\cmark}{\ding{51}}%
\newcommand{\xmark}{\ding{55}}%

\newcommand{\tCode}[1]{\texttt{#1}}

\newcommand{\meanSD}[2]{$(M\!=\!#1,~SD\!=\!#2)$}
\newcommand{\pValue}[1]{$p\!=\!#1$}

\newcommand{\anova}[5]{$(\mathrm{#1}~F(#2,#3)\!=\!#4,~p\!=\!#5)$}

\newcommand{\DSCF}[2]{$(W\!=\!#1,~p\!=\!#2)$}

\newcommand{\ChiSquare}[3]{$(\chi^2(#1)\!=\!#2,~p\!=\!#3)$}
\newcommand{\ChiSquarePL}[3]{$(\chi^2(#1)\!=\!#2,~p\!<\!.001)$}

\newcommand{\corrPL}[2]{$r(#1)\!=\!#2,~p\!<\!.001$}

\newcommand{\studentT}[3]{$t(#1)\!=\!#2,~p\!=\!#3$}
\newcommand{\yuenT}[3]{$\mathrm{Yuen's-}t(#1)\!=\!#2,~p\!=\!#3$}

\newcommand{\mannWhitney}[2]{$U\!=\!#1,~p\!=\!#2$}

\newcommand{\kmo}[1]{$(\mathrm{KMO}\!=\!#1)$}

\begin{document}
\maketitle

\begin{abstract}
	Advancements in technology are steering attention toward creating comfortable and acceptable driving characteristics in autonomous vehicles.
Ensuring a safe and comfortable ride experience is vital for the widespread adoption of autonomous vehicles, as mismatches in driving styles between humans and autonomous systems can impact passenger confidence.
Current driving functions have fixed parameters, and there is no universally agreed-upon driving style for autonomous vehicles.
Integrating driving style preferences into automated vehicles may enhance acceptance and reduce uncertainty, expediting their adoption.
A controlled vehicle study $(N=62)$ was conducted with a variety of German participants to identify the individual lateral driving behavior of human drivers, specifically emphasizing rural roads.
We introduce novel indicators for assessing stationary and transient curve negotiation, directly applicable in developing personalized lateral driving functions.
To assess the predictability of these indicators using self-reports, we introduce the MDSI-DE, the German version of the Multidimensional Driving Style Inventory.
The correlation analysis between MDSI factor scores and proposed indicators showed modest but significant associations, primarily with acceleration and jerk statistics while the in-depth lateral driving behavior turned out to be highly driver-heterogeneous.
The dataset including the anonymized socio-demographics and questionnaire responses, the raw vehicle measurements including labels, and the derived driving behavior indicators are publicly available at \href{https://www.kaggle.com/datasets/jhaselberger/spodb-subject-study-of-lateral-vehicle-guidance}{\textbf{https://www.kaggle.com/datasets/jhaselberger/spodb-subject-study-of-lateral-vehicle-guidance}}.

\end{abstract}


\section{Introduction}

The transition to complete autonomous driving remains a lengthy process \cite{roos2020technologie}.
Although \ac{avs} have been a long-standing subject of research \cite{Cox1991}, they are still being actively developed due to rapid improvements in hardware and software technology \cite{eltawab2020,Yan2020}.
With increasing technical possibilities, the focus is shifting from pure feasibility to the implementation of acceptable and comfortable driving characteristics \cite{bellem2018comfort}.
In terms of autonomous driving and advanced assistance systems, people may not readily adopt a new technology solely because it exists \cite{wintersberger2016automated}.
Systems should not only fulfill their explicit promises but also the expectations that users have \cite{drewitz2020towards}.
A comfortable and safe ride experience is crucial for the adoption of AVs \cite{bellem2018comfort,lee2021assessing,voss2018investigation,xiao2010comprehensive,bengtsson2001adaptive}.
Although driving may be technically safe, the passenger may not feel secure due to the mismatch in driving styles between the human and the autonomous system \cite{bolduc2019multimodel}.
Hence, driving styles play a crucial role in fostering trust and acceptance of AVs \cite{ekman2019exploring,strauch2019real,carsten2019can,ramm2014first}.
Up to now, there is no comprehensive and standardized definition for the classification of driving styles \cite{chu2020self,chen2021driving} and various definitions of the term "driving style" exist, as scientists operate within different conceptual or mathematical frameworks \cite{itkonen2020characterisation}.
A commonality in all definitions is that driving style encompasses a collection of driving habits that a driver develops over time as their driving experience grows \cite{elander1993behavioral, lajunen2011self,sagberg2015review,kleisen2011relationship,tement2022assessment}.
The driving style is a relatively stable driver characteristic \cite{saad2004behavioural,sagberg2015review,he2022individual} that does not undergo significant changes within a short time period, as it is a reflection of deeply ingrained driving habits \cite{martinez2017driving,tement2022assessment}.
The driving style is independent of the driving skill \cite{lajunen2011self}, as the underlying habits do not necessarily change with increasing experience \cite{tement2022assessment}. Driving skill encompasses all information-processing and psychomotor abilities of a driver.
Current driver models or driving functions represent an averaged driver with fixed parameters \cite{hasenjager2019survey, chu2020self,gao2020personalized} with no adaptation to a particular driver \cite{ponomarev2019adaptation,rosenfeld2012learning,rosenfeld2012towards,karlsson2021encoding}.
The predetermined driving style parameters provided by the manufacturer may not be comfortable for all users \cite{bolduc2019multimodel}.
The German Ethics Commission on Automated and Connected Driving criticized, that currently, the user has to adapt to the machine and not vice versa \cite{difabio2017bericht}.
The utilization of driving style preferences in automated vehicles is a rather new topic \cite{hasenjager2019survey} and challenging, as most drivers do not know their preferred parameters themself \cite{bruck2021investigation}.
It remains unclear how drivers prefer to be driven in autonomous vehicles \cite{rossner2019diskomfort, rossner2022also, gasser2013herausforderung,radlmayr2015literaturanalyse}.
However, there is substantial evidence that an adaptation towards the human driver could lead to an improved and faster acceptance and reduction of uncertainty \cite{martinez2017driving, sun2017research, bruck2021investigation, drewitz2020towards,chen2019driving,pion2012fingerprint,van2018relation,sun2020intention,gkartzonikas2019have,buyukyildiz2017identification,inagaki2003adaptive,bar2011probabilistic,chu2017curve,karlsson2021encoding,phinnemore2021happy}.
It is assumed that factors influencing perceived safety in manual driving also apply to perceived safety during automated driving \cite{rossner2020care}.
Based on the assumption and supported by scientific evidence, there is currently a prevailing inclination for drivers to prefer a driving style that aligns with their own \cite{hasenjager2019survey,festner2016einfluss,griesche2016should,bolduc2019multimodel,sun2020exploring,hartwich2015drive,rossner2022also, dettmann2021comfort,natarajan2022toward}.
While previous works primarily focus on the longitudinal aspects of driving style, the lateral component, especially on rural roads, significantly contributes to road safety.
Frequently, run-off-the-road crashes and near-crash incidents can be attributed to drivers' poor lane-keeping performance \cite{ghasemzadeh2018utilizing,ghasemzadeh2017drivers}.
Utilizing \ac{ldw} systems can lead to a potential reduction of up to 70 percent in the likelihood of an accident caused by a lane departure \cite{xu2015new} but only if they stay actively enabled \cite{johnson2016driver}. 
If such warning systems are not tailored to the driver, false alarms or missed alarms may occur \cite{deng2019curve,johnson2016driver}.
Given the high rate of false alarms, LDW systems are not frequently engaged \cite{braitman2010volvo, eichelberger2014volvo, reagan2018crash}.
Also, there is often a mismatch between the dynamic trajectory of the human driver and the more rigid driving policy of current \ac{lks}.
These systems predominantly adopt a strict center-lane following approach, which, however, is relatively uncommon in real-world driving scenarios \cite{gordon2014modeling,rossner2020care,barendswaard2019classification}.
Additionally, trajectories derived from a human driving style exhibit substantial potential to enhance perceived safety \cite{bellem2017can,lex2017objektive,rossner2018drive,rossner2019you}.

According to \cite{sagberg2015review} data collection methods for evaluating driving style can be classified into two categories: behavior recordings and self-reports. 
For the real-world driving experiment, we utilize our JUPITER research vehicle \cite{haselberger2022jupiter}. The platform features near-series placement of three additional LIDAR sensors, a precise localization system, and the use of the built-in camera and actuators.
In terms of the subjective scales, numerous self-report measures assessing driving behavior, style, and cognition have been developed and validated over the past few decades.
Frequently utilized scales encompass the \ac{dbq} \cite{reason1990errors}, the \ac{dbi} \cite{gulian1989dimensions},  the \ac{dsq} \cite{french1993decision}, and the \ac{dvq} \cite{wiesenthal2000driving}.
The \ac{dbq} is among the most widely utilized self-report assessments for driving behaviors, yet it does not account for positive behaviors \cite{rengifo2021driving}.
The \ac{dbq} evaluates deviant driving behaviors, including speeding, traffic violations, and distractions. The \ac{dbq}'s widespread popularity is attributed to its close correlation with accident involvement \cite{deng2019curve}.
Nevertheless, these scales usually focus on just one or two aspects of driving style \cite{xu2023consistency}. 
The \ac{mdsi} \cite{taubman2004multidimensional} adapted items from multiple other questionaries including the \ac{dbi}, \ac{dbq}, Driver Behavior Questionnaire \cite{furnham1993personality}, and \ac{dsq}.
The inventory consists of 44 items and encompasses four driving style aspects: patient and careful, angry and hostile, reckless and careless, and anxious.
There are translations and validations of the scale for application in various cultural contexts \cite{freuli2020cross, poo2013reliability, wang2018effect, long2019reliability, holman2015romanian, van2015measuring, padilla2020adaptation}.
Self-reports have proven to be reliable and valid \cite{taubman2016value,lajunen2003can} and the \ac{mdsi} was found to be a good indicator of driving behavior \cite{kaye2018comparison,van2018relation,taubman2016value}.
However, results obtained from self-reports may not always be ideal for objective classifications \cite{kovaceva2020identification}, as they can be influenced by a tendency to provide socially desirable responses \cite{crowne1960new,lajunen1997speed,bakhshi2022evaluating,tement2022assessment}. 
Therefore, the degree to which self-report measures accurately reflect actual driving behavior may be questioned at times \cite{af2017driver, af2015reliable,evans2004traffic,helman2015validation}.

This study emphasizes the lateral driving aspect on rural roads to identify indicators of driving behavior that objectively represent various human driving styles.
Socio-demographic dependencies, as well as the agreement of subjective self-assessments with the collected objective indicators, are examined to evaluate their usability for predicting driving style preferences for automated vehicles. 
Our contributions can be summarized to:
\begin{enumerate}
    \item Conduction of a controlled driving study utilizing a research vehicle equipped with a full multimodal sensor set.
    \item Translation of the MDSI into German resulting in the MDSI-DE including additional items.
    \item Introduction of novel driving style indicators for evaluating the lateral driving behavior.
    \item Statistical evaluation of subjects' individual driving characteristics and self-assessments.
    \item Evaluation of the agreement of subjective self-assessments with the collected objective indicators.
    \item Publicly accessible provision of the dataset including the anonymized socio-demographics and questionnaires responses, the raw vehicle measurements including labels, and the derived driving behavior indicators.
\end{enumerate}
\section{Related Work}
The socio-demographic factors age, gender, driving experience, and annual mileage are found to correlate with unsafe driving behaviors \cite{french1993decision, boyce2002instrumented, roidl2013introducing, waylen2001passenger, shinar2001self} and are also expected to have an influencing factor on a person's driving style \cite{chen2019driving,sagberg2015review}.
Particularly for gender and age, certain driving behavior indicators demonstrated significant correlations \cite{van2018relation}.
While male drivers tend to exhibit a higher propensity for exceeding speed limits \cite{chen2019driving,french1993decision,roidl2013introducing,waylen2001passenger,shinar2001self},  
female drivers demonstrate higher variability in speed \cite{van2018relation}.
In addition, male drivers exhibited an increased tendency for higher lateral accelerations \cite{roidl2013introducing} and stronger decelerations \cite{van2018relation}.
Male drivers also exhibited significantly more negative jerks than female drivers \cite{kovaceva2020identification}.
Maladaptive driving styles decline with increasing age \cite{taubman2004multidimensional}.
Previews studies show, that younger drivers tend to show higher mean speed values \cite{boyce2002instrumented,farah2011age} and an overall reduced anticipatory behavior \cite{pion2012fingerprint}.
Younger drivers often take higher risks due to their relative inexperience compared to middle-aged and older drivers \cite{strayer2004profiles}, therefore increasing age is associated with a decrease in average speed \cite{starkey2016role,taubman2004multidimensional}.
Regarding the self-evaluations, among older drivers, there is a tendency for higher scores in the careful driving style and decreased scores in the anxious, dissociative, and distress-reduction driving styles of the MDSI \cite{van2018relation,starkey2016role,taubman2004multidimensional,sagberg2015review,poo2013reliability,taubman2013family}.
Women tend to exhibit a more dissociative, anxious, and patient driving style while men scored higher on high-velocity, risky, and angry driving styles \cite{taubman2016value,poo2013study,taubman2004multidimensional,holland2010differential,poo2013reliability,wang2018effect,taubman2016multidimensional,long2019reliability,starkey2016role,gwyther2012effect}.
This is in line with the finding, that women are less likely to express driving aggression \cite{bakhshi2022evaluating,parishad2020validation}.

The MDSI has been found to be a good indicator of driving behavior in several previous studies \cite{kaye2018comparison,van2018relation,taubman2016value, zhao2012investigation,lajunen2003can,rengifo2021driving}. 
In \cite{taubman2016value} all four driving behavior indicators, consisting of the headway distance, speed, time spent over the speed limit, and time driving in the opposite lane, showed a significant correlation with scores on the reckless driving scale. In addition, the number of risky events exhibited a positive correlation with reckless and careless driving styles and a negative correlation with anxious and careful driving styles. 
In a simulator study, \cite{van2018relation} significant yet modest correlations were observed between self-reported careful driving style and speed, the average and standard deviation of the lateral position, and headway distance. Additionally, significant correlations between self-reported risky driving and the average speed and standard deviation of the lateral position were found. The scores on the angry driving style exhibit modest but significant correlations with the speed and headway distances on the highway. Also utilizing a driving simulator, \cite{rengifo2021driving} discovered that subjects exhibiting higher scores on the hostile driving style scale also demonstrated increased average speed and jerk values. In \cite{farah2009passing} the scores for angry and hostile driving styles significantly influence critical passing gaps. 
Similar results were found using the DBQ questionnaire in \cite{zhao2012investigation}. In contrast to drivers with low violation scores, those with high violation scores showed a tendency towards faster driving with more abrupt accelerations. High violation scores also correlated with larger standard deviations of steering wheel angle, increased lane changes, and more time spent in the left lane. Moreover, drivers with elevated lapse scores demonstrated greater variations in velocity, more rapid throttle accelerations, and increased frequency of steering wheel reversals. 
This is in line with the findings in \cite{helman2015validation}, where the violation subscale exhibits a moderate and significant correlation with the measured speed values in an instrumented vehicle and a simulator. Also in \cite{deng2019curve}, both the violations subscale and the errors subscale demonstrated significant correlations with the ratio of drivers' actual selected speed on curves to the theoretical curve speed.
The overall found correlations are consistent with the general evidence, that an aggressive driving style is typically linked to higher speeds, sudden accelerations and decelerations, abrupt steering changes, and harsh lateral and longitudinal maneuvers \cite{aljaafreh2012driving,castignani2015driver,deng2017driving,wang2017driving}. Calm driving styles are linked with lower speeds, gradual acceleration and braking, smooth lateral and longitudinal maneuvers, and minimal changes in steering wheel positions \cite{castignani2015driver,deng2017driving}. Normal driving styles typically fall between the spectrum of aggressive and calm driving \cite{deng2017driving,wang2017driving,aljaafreh2012driving}.

The relevant distinctive attributes of related driving studies are summarized in \autoref{tab:relatedwork}.
In terms of the research focus, it can be summarized, that previous works targeted predominantly the longitudinal aspect of human driving on highways.
In particular, when considering driving on rural roads, there is a shortcoming in the in-depth analysis of curve-negotiation behavior, as often only a basic statistical evaluation of acceleration profiles is conducted.
Context awareness is particularly vital in this context to systematically incorporate interfering influences, such as those from traffic or weather, into the evaluation.
There are numerous references to the fact that the driving context influences the driving style \cite{shouno2018deep, bellem2018comfort, bejani2018context, xu2015establishing,xue2019rapid,ericsson2000variability,chen2021driving,liu2021exploiting,ossig2022tactical}.
The driving situation offers opportunities or imposes constraints on action selection \cite{sagberg2015review}.
External factors prevent drivers from achieving theoretically ideal acceleration profiles \cite{magana2018method}.
In this regard, the extent of data collection and the selection and quantity of sensors used are also crucial for building an understanding of the driving situation before or during data analysis.
While the sample size is roughly in the same ballpark, there is a larger variance in the temporal and spatial scope of the studies, reaching from an analysis of only a few curves \cite{deng2019curve} to a naturalistic driving study over several months \cite{taubman2016value}.
When it comes to the complexity of the chosen driving behavior indicators, the majority of previous works rely on simple statistics derived from the recorded driving data. For lateral driving, these methods do not directly distinguish or categorize different trajectories or curve-cutting styles, limiting their suitability for trajectory-driven driving functions. 
In \cite{barendswaard2019classification} an evaluation of a rule-based trajectory classifier capable of distinguishing seven curve-negotiation styles is conducted through a simulator subject study. The study identified curve cutting and biased inner curve negotiation as the most common for right turns, and biased outer curve negotiation and curve cutting as the most common for left turns. Nevertheless, the study's ability to generalize to real driving situations is limited due to the use of a relatively simple driving simulator and trajectories were recorded only on three artificial curves with a constant velocity profile.
The scope of the related work is evident in the choice of questionnaires used. Many studies focused on evaluating angry, risky, or reckless driving, with the \ac{dbq} being one of the most commonly employed inventories. Occasionally the \ac{pss} \cite{cohen1983global}, \ac{deq} \cite{amado2014accurately}, \ac{dseq} \cite{amado2014accurately}, \ac{rdhs} \cite{taubman2004multi}, \ac{cas} \cite{lee2020coronavirus}, 
\ac{aq} \cite{buss1992aggression}, and
\ac{das} \cite{deffenbacher1994development} is used.

To the authors' knowledge, we are the first to make our dataset directly publicly accessible including the anonymized socio-demographics and questionnaire responses, the raw vehicle measurements including labels for road types, street names, oncoming traffic, vehicle-following situations, and the derived driving behavior indicators.
We aim to further facilitate research on driving style and driving behavior analysis not only for lateral but also for longitudinal driving.
Only a few other studies retain the option to share the data on request; however, there is no information about the extent of the data.

\begingroup

\setlength{\tabcolsep}{2pt} 

\begin{table}[]
    \caption{Related subject studies in chronological order. N represents the number of subjects, the country codes follow the ISO 3166-1 encoding. \cmark, (\cmark), and \xmark~indicate whether a requirement is met, partially met or not fulfilled. (-/+/++/+++) are denoting, as objectively as possible, the richness of the recorded objective data: - for no sensor data, + for time series data only, ++ for time series and additional sensors like cameras, and +++ for a full multimodal sensor set. For the KPI complexity: - denotes no KPI based on objective data, + indicates basic statistics of driving data, ++ represents relatively simple derived indicators or model parameters, and +++ signifies complex KPIs.}
	\centering
    \scriptsize
    \begin{tabular}{lccclcclllllcclccccl}
    \toprule
    \textbf{}               & \textbf{}     & \textbf{}  & \textbf{}            & \textbf{} & \multicolumn{2}{c}{\textbf{Domain}} & \textbf{} & \multicolumn{3}{c}{\textbf{Context}}              & \textbf{} & \multicolumn{2}{c}{\textbf{Focus}}       & \textbf{} & \textbf{}                   & \textbf{}                  & \textbf{}               & \textbf{}                              & \textbf{}                  \\ \cline{6-7} \cline{9-11} \cline{13-14}
    \textbf{Ref}            & \textbf{Year} & \textbf{N} & \textbf{Country}     & \textbf{} & \rotatebox{90}{\textbf{Real}} & \rotatebox{90}{\textbf{Simulation}} & \textbf{} & \rotatebox{90}{\textbf{City}} & \rotatebox{90}{\textbf{Rural}} & \rotatebox{90}{\textbf{Highway}} & \textbf{} & \rotatebox{90}{\textbf{Lateral}} & \rotatebox{90}{\textbf{Longitudinal}} & \textbf{} & \rotatebox{90}{\textbf{Data Capture Level}} & \rotatebox{90}{\textbf{Context Awareness}} & \rotatebox{90}{\textbf{KPI Complexity}} & \textbf{Questionnaires}                 & \rotatebox{90}{\textbf{Data Availability}} \\
    \midrule
    \cite{farah2009passing}        & 2009          & 100        & IL & & &  \cmark & & & \cmark & & & & \cmark & & + & \cmark & ++ & MDSI & \xmark  \\
    \cite{zhao2012investigation}   & 2012          & 108        & US                   &           & \cmark             &                     &           &               &                & \cmark                &           & \cmark                & \cmark                     &           & +++                       & \xmark                          & +                     & DBQ, PSS      & \xmark                          \\
    \cite{hong2014smartphone}      & 2014          & 22         & US                   &           & \cmark             &                     &           & \cmark             & \cmark              & \cmark                &           & (\cmark)              & \cmark                     &           & +                         & \xmark                        & +                     & DBQ                         & \xmark                          \\
    \cite{amado2014accurately}     & 2014          & 158        & TR               &           & \cmark             &                     &           & \cmark             & \cmark              & \cmark                &           & \cmark                & \cmark                     &           & -                         &   \xmark                          & -                     & DBQ, DEQ, DSEQ                         & \xmark                          \\
    \cite{helman2015validation}    & 2015          & 58         & GB                   &           & \cmark             & \cmark                   &           &               & \cmark              &                  &           &                  & \cmark                     &           & ++                        & \cmark                          & +                     & DBQ                                    & \xmark                          \\
    \cite{bellem2016objective}     & 2016          & 24         & DE              &           & \cmark             &                     &           & \cmark             & \cmark              & \cmark                &           & \cmark                & \cmark                     &           & ++                        & \cmark                          & +                     &                                       & \xmark                          \\
    \cite{taubman2016value}        & 2016          & 231        & IL    & &   \cmark & \cmark & & \cmark & \cmark & \cmark & & (\cmark) & \cmark & & + & \cmark & ++ & MDSI, RDHS & \xmark \\
    \cite{pariota2017validation}   & 2017          & 100        & IT                &           & \cmark             & \cmark                   &           &               & \cmark              & \cmark                &           &                  & \cmark                     &           & +++                       & \cmark                          & +                     &                                       & \xmark                          \\
    \cite{van2018relation}         & 2018          & 88         & NL, BE &           &               & \cmark                   &           & \cmark             &                & \cmark                &           & \cmark                & \cmark                     &           & +                         & \cmark                          & +                     & MDSI                                   & \xmark                          \\
    \cite{deng2019curve}           & 2019          & 24         & CN                &           & \cmark             &                     &           &               & \cmark              & \cmark                &           & (\cmark)              & \cmark                     &           & +                         & \cmark                          & ++                    & DBQ                                    & \xmark                          \\
    \cite{ramezani2021comparing}   & 2021          & 156        & IR                 &           &               & \cmark                   &           &               &                & \cmark                &           &                  & \cmark                     &           & +                         & \cmark                          & +++                   &     own              & \cmark$^\dagger$                 \\
    \cite{corcoba2021covid}        & 2021          & 30         & ES                &           & \cmark             &                     &           &               &                & \cmark                &           &                  & \cmark                     &           & +                         &    \xmark                         & +                     & CAS & \xmark                          \\    \cite{rengifo2021driving}      & 2021          & 20         & FR               &           &               & \cmark                   &           & \cmark             & \cmark              & \cmark                &           & \cmark                & \cmark                     &           & +                         & \cmark                          & +                     & DBQ, MDSI                             & \xmark                          \\
    \cite{xu2023consistency}       & 2023          & 32         & CN                &           &               & \cmark                   &           &               &                & \cmark                &           & \cmark                & \cmark                     &           & +                         & \cmark                          & +                    & MDSI                                   & \xmark                          \\
    \cite{adavikottu2023modelling} & 2023          & 58         & IN                &           &               & \cmark                   &           &               & \cmark              &               &           &                  & \cmark                     &           & +                         & \cmark                          & +                     & DAS, AQ, ADS                         & \cmark$^\dagger$                \\
    \textbf{ours}           & 2023          & 62         & DE              &           & \cmark             &                     &           & \cmark$^\ddag$            & \cmark~~              & \cmark$^\ddag$                &           & \cmark                & \cmark$^\ddag$                 &           & +++                       & \cmark                          & +++                   & MDSI                                   & \cmark          \\              
    \bottomrule
    \multicolumn{19}{l}{\footnotesize{$^\ddag$ while all longitudinal relevant data is recorded, the focus of the study lies on the lateral evaluation}} \\
    \multicolumn{19}{l}{\footnotesize{$^\dagger$ available only on request}} \\
    \end{tabular}
    \label{tab:relatedwork}
    \end{table}

    \endgroup

\section{Method}

\subsection{Instruments}
The participants received the online questionnaire prior to the driving experiment. The questionnaire is composed of two parts.
The first part consists of general questions about the subjects, covering details such as age, gender, annual mileage, frequency of car usage, possession of a driver's license, preferred average highway speed, engine power of their own car, and a one item self-assessment of driving style on highways and rural roads.
For the more sophisticated driving style inventory, the latter part consists of all items from the original version of the MDSI \cite{taubman2004multi}. Supplementary items relevant to the Angry, Risky, Careful, and Distress-Reduction Driving Style factors are incorporated from \cite{poo2013reliability} which fit into the context of rural road driving. In this regard, also the three own defined questions "If time permits, I prefer to drive on the rural road instead of the highway", "Driving on narrow rural roads overwhelms me", and "I enjoy the variation on curvy roads" are added. In summary, the questionnaire contains 57 items and is referenced hereafter as the MDSI-DE. Participants scored their agreement on each MDSI item using a 6-point Likert scale, ranging from 1 (not at all) to 6 (very much). 

To the best of our knowledge, there existed no published German version of the MDSI prior to this study.
For the translation process of the MDSI-DE we followed the approach of \cite{holman2015romanian}.
In the initial stage, two translators, fluent in English and native in German, independently performed forward translations of the 57 English items.
After resolving the differences between the two versions, the initial German version was then back-translated into English by a third professional translator, who was blind to the original English version. In the last step, the initial German version, its back-translation, and the original English documents are compared, ensuring that the final German version maintains the highest level of relevant equivalences.
The finalized versions of the MDSI-DE in both English and German are presented in \autoref{tab:participants052}, the reported values of internal consistency (Cronbach's alpha) are specific to the current sample.
While this study exclusively utilizes items from the validated MDSI version by \cite{van2015measuring} for factor score calculation and subjects' driving style estimation, the supplementary items will be employed in subsequent research studies.
\subsection{Participants}
A total of 62 participants from the general population of German drivers took part in the online survey and the measurements in the real vehicle. Participants were required to hold a driver's license for a minimum of 2 years.
In this study participants were given the freedom to self-identify their gender during the online pre-survey, choosing from options such as female, male, or diverse.
They were also given the option to provide an open response or choose not to disclose their gender.
The analysis revealed a division, with 17 female drivers (27.4\%) and 45 male drivers (72.6\%), within this sample of subjects no other gender types were specified.
The age ranged from 20 up to 70 years with an average of 33.1 $\pm$ 12.7 years. Subjects reported an average driver's license ownership of 15.9 $\pm$ 12.4 years, with a maximum value of 53 and a minimum value of 3 years. See \autoref{tab:participants} for gender-specific values.
In terms of frequency of car usage, 30 people replied that they drove daily, 23 several times a week, five weekly, and four used the car only a few times a month. The majority has an annual mileage of 10000 to 20000 kilometers. The specific gender distribution is given in \autoref{tab:participants02} in the Appendix. The question of having driven an SUV before was answered in the affirmative by 82\% of women and 77\% of men. All subjects who completed the online questionnaire also participated in the actual driving experiment. Participation for the two-fold survey study was purely voluntary, any compensation or reward was excluded. All personal data were anonymized.

\begingroup

\setlength{\tabcolsep}{3pt} 

\begin{table}[]
    \caption{\small{Age, license possession, preferred average speed on highways and engine power of personal car split into women (N=17) and men (N=45).}}
    \centering
    \scriptsize
    \begin{tabular}{llllll}
    \toprule
                           & Gender & Mean  & SD   & Min   & Max  \\ 
   \midrule
    Age (years)            & female & 32.1  & 11.2 & 20    & 58   \\
                           & male   & 33.5  & 13.4 & 21    & 70   \\
    Licence posession (years)      & female & 14.5  & 10.6 & 3  & 40 \\
                           & male   & 16.4  & 31.1 & 3.75  & 53 \\
    Travel speed highway (kmh)   & female & 150.6 & 26.8 & 100   & 210  \\
                           & male   & 145.1 & 13.4 & 125   & 180  \\
    Engine power (PS) & female & 114.7 & 51.0 & 60    & 270  \\
                           & male   & 150.2 & 62.9 & 60    & 320  \\ \bottomrule
    \end{tabular}
    \label{tab:participants}
    \end{table}

    \endgroup

\subsection{Procedure}
\begin{figure}[]
    \centering
    \includegraphics[page=1,trim={0 0cm 0 0}, clip,width=0.58\linewidth]{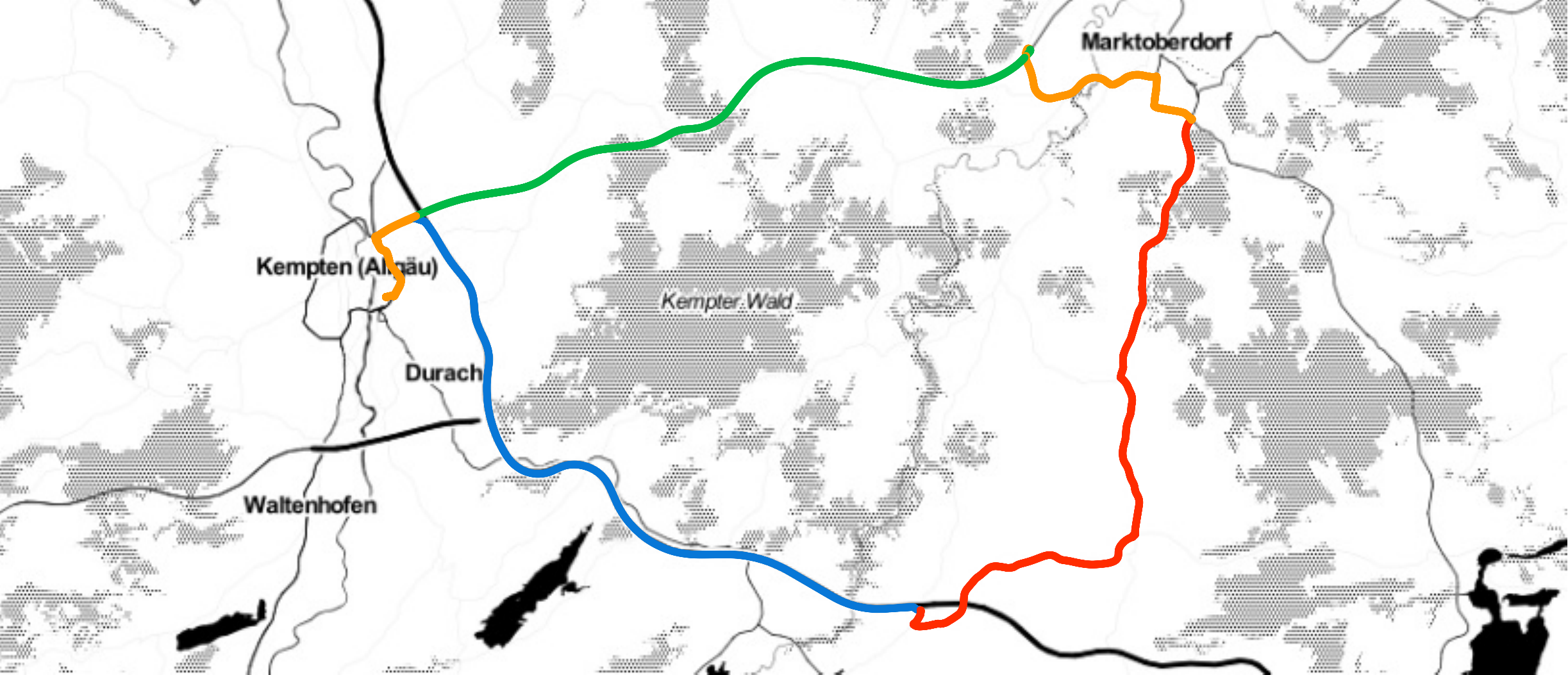}
    \caption{The used measurement track with a total length of \SI{71.9}{\kilo\metre} is located near Kempten (Allgäu), Germany and is composed of \SI{14.5}{\kilo\metre} city (orange), \SI{20.5}{\kilo\metre} highway (blue), \SI{18.3}{\kilo\metre} rural (red), and \SI{18.6}{\kilo\metre} federal (green) road segments. Each subject followed the exact same route. The track is followed counterclockwise.}
    \label{fig:route}
\end{figure}

After registering for the driving experiment, subjects were given access to questionnaires and asked to complete them in advance.
Upon arrival at the laboratory, participants signed an institutionally approved consent form, which incorporated the permission to release the anonymized data.
A research associate introduced the general aim of the study and the research vehicle without providing details about the recorded driving data and scenarios to avoid influencing the subjects.
Driving was completed in a mid-sized sports utility vehicle (Porsche Cayenne), our JUPITER platform \cite{haselberger2022jupiter}.
While equipped with a diverse multimodal sensor set from the subject's perspective, the vehicle is not distinguishable from standard production vehicles.
To allow inter-subject comparability, all subjects drove on the same route.
The circuit starts and ends in Kempten (Allgäu), Germany and consists of city, highway, rural, and federal road sections.
The exact route layout and course are shown in \autoref{fig:route}.
To ease orientation for the test persons, the vehicle's own navigation system is switched on and guidance is provided by the co-driver at defined locations.
However, no instructions take place beyond pure navigation.
The participants were instructed to drive in their usual, natural manner, with the freedom to choose their speed, lane, lane-change execution, and cornering behavior.
The familiarization period lasted approximately ten minutes.
During this phase, the driver had full freedom to adapt to the vehicle, and the data collected during this time was not utilized for the analysis.
The total distance of the experiment measures \SI{71.9}{\kilo\metre}, resulting in an average travel time of one hour.
The described work adheres to The Code of Ethics of the World Medical Association (Declaration of Helsinki) \cite{world2013world}.
Throughout the entire experiment, the road traffic regulations were strictly followed to ensure the safety of the participants and other road users.

\subsection{Objective Driving Behavior Indicators}

To evaluate the actual measurable driving behavior of the subjects, objective driving behavior indicators are derived from the recorded raw sensor data in conjunction with lane data derived from the camera.
Situations with lane changes and with detected oncoming traffic are not considered and filtered out.

\subsubsection*{Basic Indicators}
Among these indicators are the longitudinal and lateral accelerations $a_x$ and $a_y$, the distance to lane center $d_{\mathrm{CL}}$, the relative drift velocity $v_{\mathrm{drift}}$, and the longitudinal and lateral jerk values $k_x$ and $k_y$. The distance to the center of the lane is determined by the distances to the left and right lane boundaries of the ego lane. The relative drift velocity is indicated in percent and given as:
\begin{equation}
    v_{\mathrm{drift}} = \frac{\delta d_{\mathrm{CL}}}{\delta t}\frac{100\%}{v_x}
\label{eq_rel_drift}
\end{equation}
Jerk values describe the rate of change of accelerations and thus represent how smoothly a person drives. Jerk values are also directly linked to perceived ride comfort \cite{bae2020self}.
In the context of driving behavior analyses, jerk values have proven themselves in the classification of driving styles and are therefore frequently used \cite{van2018relation, murphey2009driver, rath2019lane, feng2018driving}.

\subsubsection*{G-G Envelope}
\begin{figure}[]
    \centering
    \includegraphics[page=1,trim={0 1.6cm 0 0}, clip,width=0.45\linewidth]{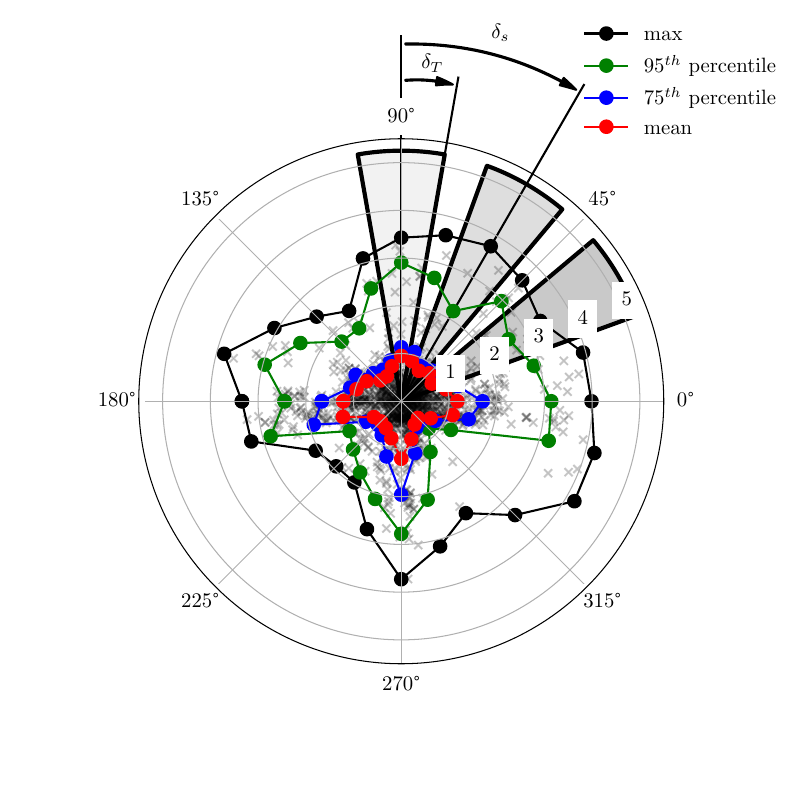}
    \caption{Determination of the G-G Envelope points. Small angle pieces of size $2\delta_r$ are rotated using $\delta_s$ to sample the acceleration points. The statistical features maximum, 95$^{th}$ percentile, 75$^{th}$ percentile, and mean are calculated for each angle bin to form the envelope. In the background, the raw measurements are shown.}
    \label{fig:ggEnvelope}
\end{figure}

G-G diagrams are used to combine the longitudinal and lateral accelerations into a representation of the driver's driving style preferences and their perceived level of risk associated with the dynamic movement of the vehicle and can be integrated into vehicle control systems \cite{kalabic2019learning,bae2020self}. In addition, the comfort limits of the drivers can be identified. For example, it can be observed that the normal driver perceives lateral accelerations greater than $4 \frac{m}{s^2}$ as unpleasant at driving speeds above $50$ km/h and therefore avoids them \cite{pion2012fingerprint}. The utilization of dynamic reserves in the diagram differs from driver to driver \cite{schacher2018fahrerspezifische}. Unlike the normal driver, experienced racing drivers take full advantage of the performance envelope, as they are not concerned about vehicular instability \cite{bae2020self}. Under good road conditions, sports cars can achieve an absolute acceleration of approximately $9.81 \frac{m}{s^2}$ \cite{lopez1997going}.
The resulting shape of the G-G diagram is a relevant indicator for the driving style, as they develop quickly and remain consistent over time \cite{will2020methodological,wegschweider2005modellbasierte}. Usually, patterns in the shape of a heart, leaf or flat profile can be found. Previously, these profiles were evaluated only manually. In \cite{will2020methodological}, the diagrams are printed and clustered based on the visual observation of three experts. To overcome this limitation, we propose a novel approach to automatically analyze the G-G shapes using their envelopes. First, the longitudinal and lateral acceleration values are treated as cartesian coordinates and then converted to the polar coordinate system. The resulting accelerations are sampled by a stridden rotation with small angular intercepts. To counteract a too sparse sampling, a tolerance range of $\pm \delta_r$ around the current angle value is defined. Within these angle pieces, the statistical indicators mean, maximum, 75th percentile, and 95th percentile are derived.
Starting from pure straight driving, these pieces are rotated by $\delta_s$ until the full diagram has been covered. This results in $N_E$ envelope reference points with four KPIs each, where $N_E$ solely depends on $\delta_s$:
\begin{equation}
    N_E = \frac{2\pi}{\delta_s}
\label{eq:ggEnvelopeN}
\end{equation}
One advantage of this method is that the number of interpolation points is independent of the number of measured acceleration values. The granularity of the envelope can be freely specified by $\delta_s$, if $\delta_r > \frac{\delta_s}{2}$, the angle sections overlap. The procedure is visually represented in \autoref{fig:ggEnvelope}.
For driving behavior analysis, the $N$ envelope points can theoretically be used directly. To simplify for instance clustering, a \ac{pca} is used to reduce the dimensionality to a two-dimensional space.

\subsubsection*{Stationary Cornering Behavior}
\begin{figure}[t]
		\centering
		\small
		\begin{tabular}{ccc}
			\includegraphics[width=.28\linewidth,valign=m]{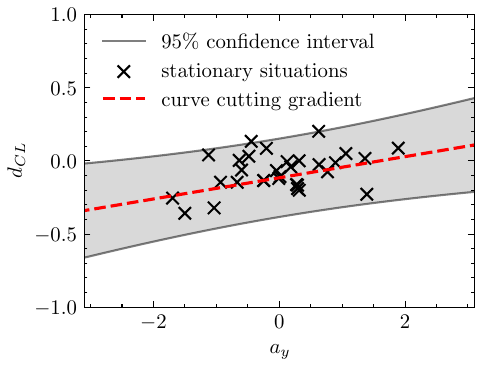} & 
			\includegraphics[width=.28\linewidth,valign=m]{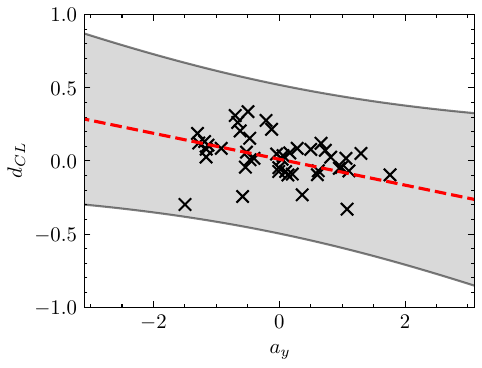} &
			\includegraphics[width=.28\linewidth,valign=m]{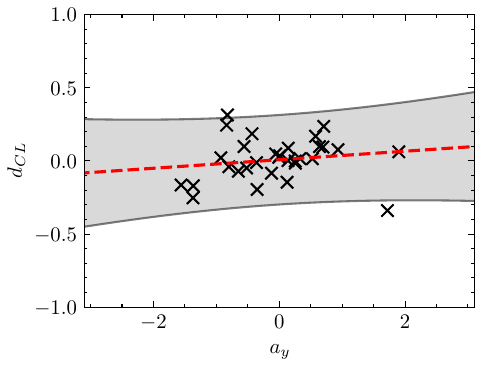} \\
		\end{tabular}
		\caption{Stationary cornering behavior on rural roads without traffic exemplified on three randomly selected subjects. The lateral acceleration $a_y$ and the distance to the center-line $d_{CL}$ are shown on the x-axis and y-axis. The curve-cutting gradient is derived from a linear regression of the stationary cornering behavior. While the first and third subjects cut the curve, the second subject tends to drift outward.}
\label{fig:ccg}
\end{figure}

To quantify the lateral deviation during cornering, in \cite{apel1998modellierung} the driving line is defined as the average lane position over the road curvature $\kappa$ under free traffic and for sections with constant curvature.
Following on from this, in \cite{laubis2020ccg} the lateral acceleration $a_y$ is used instead of the curvature to address the neglect of velocity $v$.
For constant circular motion, their relationship can be represented by:
\begin{equation}
a_y = \kappa v_x^2 
\end{equation}
To meet the constant curvature condition, quasi-stationary situations are extracted from the driving data.
\autoref{fig:ccg} exemplifies for three subjects their stationary cornering behavior for trips on rural roads without traffic.
Several objective KPIs can be derived from this representation of curve driving. 
Using linear regression, the gradient, hereafter referred to as \ac{ccg}, of the distance to the centerline $d_{CL}$ and the lateral acceleration $a_y$ can be assigned to all stationary scenarios.
A positive $\mathrm{CCG}$ indicates sportive driving, while negative values occur if the vehicle drifts to the outside of the curve. The intersection point with the y-axis is used to determine the global offset $\mathrm{CCG}_0$. In addition, the width of the 95$\%$ confidence interval $\mathrm{CI}_{d_{CL}}$ allows determining the consistency. The narrower the band, the more reproducible and accurate the driver's performance.
Based on its strong interpretability, this parameter is also applied in the industry for attribute-based development of driver assistance systems \cite{hofer2020attribute}.


\begin{figure}[t]
	\centering
	\small
	\begin{tabular}{cccc}
		\includegraphics[width=.19\linewidth,valign=m]{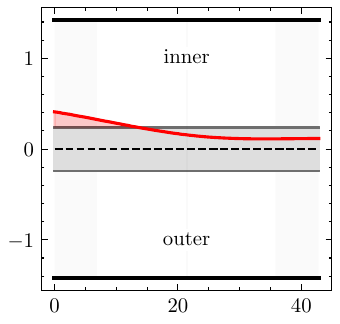} & 
		\includegraphics[width=.19\linewidth,valign=m]{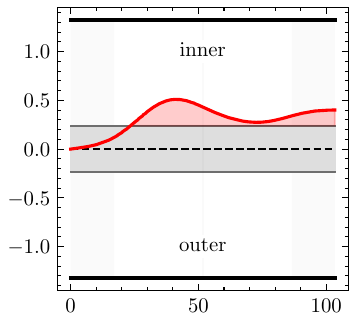} &
		\includegraphics[width=.19\linewidth,valign=m]{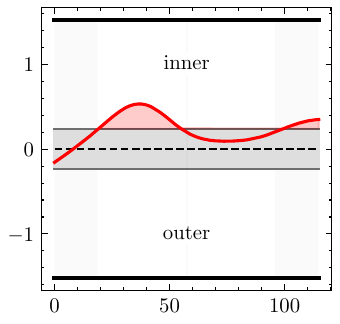} &
		\includegraphics[width=.19\linewidth,valign=m]{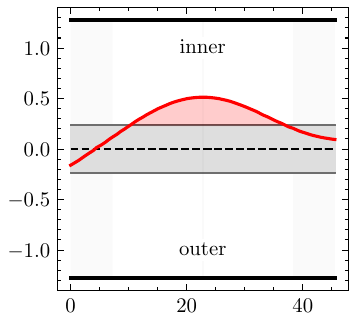} \\
		\tCode{IICC} & \tCode{CIII} & \tCode{CICI} & \tCode{CIIC} \\
		Early Cutting & Biased Inner & Oscillating Cutting & Cutting \\
		~ & ~ & ~ \\
		\includegraphics[width=.19\linewidth,valign=m]{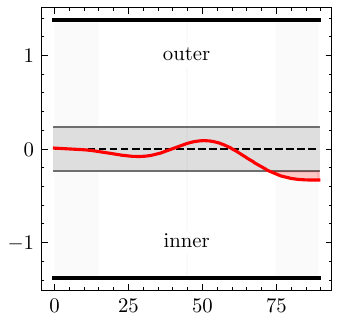} & 
		\includegraphics[width=.19\linewidth,valign=m]{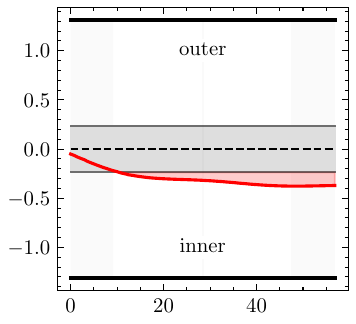} &
		\includegraphics[width=.19\linewidth,valign=m]{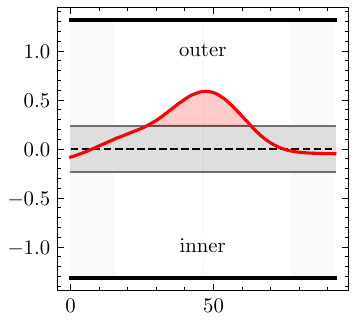} &
		\includegraphics[width=.19\linewidth,valign=m]{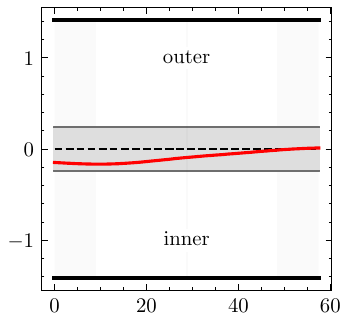} \\
		\tCode{CCCI} & \tCode{CIII} & \tCode{COOC} & \tCode{CCCC} \\
		Late Cutting & Biased Inner & Counter Cutting & Center \\
	\end{tabular}
	\caption{Examples of recorded trajectories with associated four-digit codes and derived trajectory classes. The left curves are shown in the first row, and the right curves in the second row. All curves are projected onto a straight road surface. The center band is illustrated in gray, and the theoretical lane center as a dashed line. Curve entries and exists are shaded with light gray. The driven distance is displayed on the x-axis and the lateral deviations on the y-axis. The curve-cutting intensity is visualized by the light red shaded areas. The outermost two lines represent the lane boundaries.}
\label{fig:trajectoryClassification}
\end{figure}


\subsubsection*{Transient Cornering Behavior}
While the quasi-stationary curve behavior is evaluable with the help of the CCG, in \cite{barendswaard2019classification} a rule-based classifier is proposed within a simulator subject study which also considers the transient driving behavior. Based on the recorded lateral positions relative to the road centerline, a trajectory class is derived for each curve. The method relies on an evaluation of transitions across a center band to differentiate an intentional curve cutting. As center-band tolerance the average  standard  deviation $\overline{S}_{d_{CL}}$  of  straight  lane driving is used:   
\begin{equation}
    \overline{S}_{d_{CL}} = \frac{1}{N} \sum_{i=1}^{N}S_{d_{CL},i}
\end{equation}
where $N$ is the number of subjects and $S_{d_{CL},i}$ the subject-specific straight lane driving behavior. In \cite{barendswaard2019classification} a band with \SI{\pm 0.10}{\meter} is used based on the findings of a prior simulator study \cite{scholtens2018new}. The real-world driving experiments in this work, however, lead to a wider center-band tolerance of \SI{\pm 0.24}{\meter}. This deviation can be attributed on the one hand to the non-negligible domain gap between the simplified simulation environment and the real-world road conditions and on the other hand to the used vehicle type. With a large vehicle, small deviations from the center of the lane are less noticeable.
In \cite{barendswaard2019classification} the curves are divided into three segments entry, in-curve, and exit.
The assumption is made that a maximum of two centerline transitions can occur.
However, the experiments in this work revealed that this simplification is violated for some subjects and multiple transitions exist.
To overcome this, a division into four sections is proposed, where for both the entry and exit, one-sixth of the curve is used, and one-third for each of the first and second halves.
To derive a corresponding trajectory class for each curve, it is first checked whether the 75$^{th}$ percentile of the trajectory points lay within the center band.
Accordingly, the labels Center \tCode{C}, Inner \tCode{I}, and Outer \tCode{O} are assigned to each segment to form a four-digit curve encoding. 
The final trajectory label is derived using the encoding and the class mapping which can be found in \autoref{tab:trajectoryClassMapping} in the appendix. 
In total, \num{16} trajectory classes are defined, consisting of the seven classes from \cite{barendswaard2019classification} and nine new classes derived from the measurements. Samples with the associated curve encoding and the inferred trajectory class can be seen in \autoref{fig:trajectoryClassification}.
The class percentage is calculated from the trajectory labels and normalized to the number of evaluable curves per subject.
To evaluate the degree of curve cutting, the intensity is calculated in addition to the pure classification. In this work, the intensity $I_T$ is defined as the area of the trajectory outside the center-line tolerance band $\overline{S}_{d_{CL}}$: 
\begin{equation}
    \acute{I_T} = \frac{\int_{0}^{\tau}\max(|d_{CL,t}|-\overline{S}_{d_{CL}},0)dt}{\int_{0}^{\tau}v_tdt}
\end{equation}
To account for the different lengths of the curves, the intensity is normalized to the distances traveled by the subjects.
Statistical descriptors are calculated for the intensity values separately for each trajectory class.


\subsection*{Selection of Curves of Interest}
To provide inter-subject comparability, the stationary and transient cornering characteristics are evaluated using geo-spatially labeled curves. To specify the boundaries of the curves, a curvature threshold $\tau_{\kappa}$ of \SI{0.002}{\per\metre} is used, which corresponds to a curve radius of \SI{500}{\metre}. This threshold was selected as up to this radius no change in driving behavior is observed compared to straight sections on rural roads \cite{zierke2010sichere,buck1992geschwindigkeitsverhalten}.
After applying this labeling strategy, \num{49} curves of interest are selected for rural road driving.


\section{Results}

The statistical analysis of both the subjective self-evaluations and the objective driving indicators was performed in jamovi \cite{jamovi2023jamovi}, an open statistical software.
Before conducting each individual analysis, we verified the assumptions of normal distribution and homogeneity of variance using Shapiro-Wilk and Levene's tests.
If both normality and homogeneity assumptions are violated, we employ the robust independent samples t-test and the robust \ac{anova} variant available in jamovi.
For the evaluation of the age data, the subjects were divided into three age groups: young ($< 25$), middle-aged ($25 - 54$), and old ($ > 54$).
Similar age groups were utilized in \cite{mehmood2009modeling, hartwich2018driving, chen2019driving, holman2015romanian, gwyther2012effect, padilla2020adaptation}.

\begin{figure}[t]
    \centering
    \includegraphics[page=1,trim={0 0cm 0 0}, clip,width=0.49\linewidth]{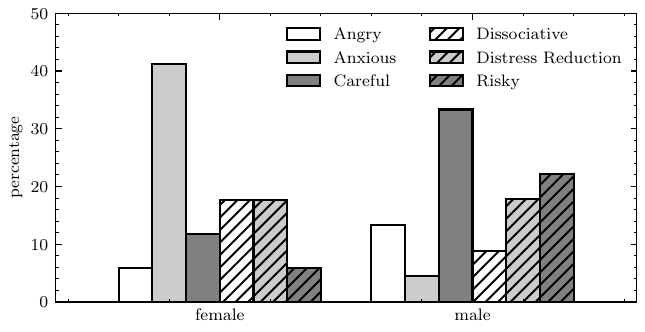}
    \caption{Distribution of driving styles determined from the MDSI questionnaire divided into women and men. Within each gender group, all six driving styles are represented.}
    \label{fig:mdsiRefindedStyleDistribution}
\end{figure}
\subsection{Driving Style Self-Assessment}
To compare the self-assessment with the objective driving characteristics, in the first step the factor scores are calculated based on the MDSI questionnaire.
Since the number of subjects within this work is considered inadequate for significant factor analysis of the MDSI items, the factor divisions and loadings are taken from \cite{van2015measuring}.
In \cite{van2015measuring} an online subject study (N=\num{364}) was conducted to validate the stability of the MDSI for the Netherlands and Belgium revealing six driving styles. 
Evidence, that the original eight driving styles of the MDSI can be reduced to these six is also reported by recent studies \cite{nees2021relationships, long2019reliability, padilla2020adaptation, trogolo2020factor, holman2015romanian}. This version of the MDSI is also applied in \cite{van2018relation}.
In this study, using the highest scoring factor as a classification feature would lead to an assignment of $87\%$ to the careful driving class which, however, does not match the subjective impressions of the co-pilots.
To overcome this, in \cite{van2015measuring, karjanto2017identification, nees2021relationships} refined scores are estimated by the multiple regression approach which maximizes the validity of the estimates \cite{distefano2009understanding}. 
This method provides factor scores that indicate each subject's deviation from a factor's mean \cite{andy2009discovering}.
A high factor score means that a person scores higher than the average of the scores of all participants within that factor.
The driving style class is determined by the factor with the highest average. A comparison of the non-refined and refined factor score statistics and the number of assigned subjects to each driving style can be found in \autoref{tab:mdsiNonAndRefinedScores}. The resulting refined, gender-specific distribution of the MDSI driving styles is visualized in \autoref{fig:mdsiRefindedStyleDistribution}.
Only the difference in mean scores of the Anxious driving style between women \meanSD{0.70}{1.32} and men \meanSD{-0.27}{0.58} is significant according to a Yuen-Welch's test, \yuenT{11.0}{2.56}{.03}.

\subsection{Objective Driving Behavior Indicators}

\subsubsection*{Basic Indicators}
To analyze the basic driving indicators, the eight statistics maximum of the absolute values (Absmax), maximum, minimum, mean, median, \ac{sd}, \ac{rms}, and \ac{idr} are applied.
In the field of driving style analysis, particularly for acceleration and jerk values, the RMS is commonly used \cite{lv2018driving, sun2020intention, rath2019personalised, sun2017research}.
The IDR is the distance between the 10$^{th}$ and 90$^{th}$ percentiles and measures the statistical dispersion.
In \cite{itkonen2020characterisation} the IDR is applied to acceleration and jerk values to cut off the corner case situations as these values are considered rare and unintentional.
The statistics of $a_x$, $a_y$, $d_{\mathrm{CL}}$, $v_{\mathrm{drift}}$, $k_x$, and $k_y$ are extracted separately for each of the selected curves and then averaged using either the mean or median.

\autoref{tab:accelKpiResults} lists all the statistics of the longitudinal and lateral accelerations. It is noticeable that for the lateral accelerations higher values result for each indicator compared to the longitudinal ones. Furthermore, the longitudinal and lateral statistics are significantly linearly correlated with each other. 
For example, the mean of the Absmax or the IDR values of $a_x$ and $a_y$ were found to be significantly positively correlated, \corrPL{60}{.652} and \corrPL{60}{.760}.
One-way ANOVAs were performed to compare the effect of the three different age groups on the acceleration scores.
A Kruskal-Wallis showed, that there was a statistically significant difference in mean Absmax values of $a_x$ between at least two groups \ChiSquare{2}{7.02}{.03}.
Dwass-Steel-Critchlow-Fligner pairwise comparisons \DSCF{-3.67}{.03} revealed that the mean value was significantly higher for the younger \meanSD{1.28}{0.15} than for the older age group \meanSD{1.12}{0.11}. 
For the standard deviation of the lateral accelerations, a one-way analysis of variance showed that the different age groups have a significant impact \anova{Welch's}{2}{20.5}{5.14}{.02}.
Post hoc analyses using the Games-Howell test indicated that the mean standard deviation was significantly higher for the younger subjects \meanSD{0.49}{0.02} than for the older ones \meanSD{0.45}{0.02} at \pValue{.01}. 
Also, the maximum values differ significantly between the age groups \anova{Fisher's}{2}{59}{3.30}{.04}.
According to Tukey’s HSD test for multiple comparisons, the means of the maximal lateral accelerations of the younger drivers \meanSD{2.56}{0.19} were significantly higher than for the older drivers \meanSD{2.35}{0.21} at \pValue{.04}.
No gender correlations were found for acceleration scores.

For the jerk values, which can be found in \autoref{tab:jerkKpiResults}, the values are laterally higher than longitudinally similar to the acceleration statistics. Furthermore, there exists a significant pairwise linear correlation between $k_x$ and $k_y$.
A Mann-Whitney test indicated that mean longitudinal jerk values were significantly higher for the male drivers \meanSD{0.016}{0.033} than for the female drivers \meanSD{0.002}{0.018}, \mannWhitney{247}{.03}. 
However, the Absmax values of $k_x$ were significantly lower for the male \meanSD{19.28}{2.07} compared to the female \meanSD{20.49}{1.50} subjects, \studentT{60}{2.21}{.03}.
In addition, the test runs with women \meanSD{-18.17}{1.36} had significantly higher braking jerk values compared to men \meanSD{-16.85}{1.90}, \studentT{60}{-2.01}{.01}. For both longitudinal and lateral jerk values, no age-group-related effects were found.

The statistics of the distance to the lane center, shown together with the relative drift velocities in \autoref{tab:d2clVDriftKpiResults}, indicate that, on average, the subjects drove with an offset toward the right side of the lane.
Comparing the mean $d_{\mathrm{CL}}$ values for pure straight-ahead driving \meanSD{-0.041}{0.24} with cornering \meanSD{-0.15}{0.05} results in an opposite pattern.
While high standard deviations occur when driving straight ahead, the average distance to the center lane is almost insignificant. 
But inside the curves, this reverses and the mean standard deviation becomes very low, while the mean $d_{\mathrm{CL}}$ increases.
Thereby according to a Mann-Whitney test, the mean shifts for the men \meanSD{-0.16}{0.05} are significantly higher compared to the women \meanSD{-0.13}{0.06}, \mannWhitney{259}{.05}.
One-way ANOVAs reported significant effects of age group membership on the $d_{\mathrm{CL}}$ indicators.
For the maximum distance to the lane center, a Kruskal-Wallis test \ChiSquare{2}{7.62}{.02} and Dwass-Steel-Critchlow-Fligner pairwise comparisons \DSCF{3.47}{.02} revealed, that the middle-aged group \meanSD{0.09}{0.04} had lower mean values than the older group \meanSD{0.14}{0.06}. The younger drivers \meanSD{0.08}{0.04} showed lower mean values than the older drivers, \DSCF{3.48}{.04}.
Also, the mean standard deviation and the IDR values differ significantly between the age groups, \anova{Fisher's}{2}{59}{5.59}{0.01} and \anova{Fisher's}{2}{59}{4.75}{0.01}. Thereby the younger ones \meanSD{0.11}{0.01}  showed lower mean SD values than the middle group \meanSD{0.13}{0.02} at \pValue{.03} and the older group \meanSD{0.13}{0.02} at \pValue{.02}. 
For the mean IDR ranges, also the younger drivers \meanSD{0.16}{0.02} showed the lowest values compared to the middle-aged drivers \meanSD{0.18}{0.03} at \pValue{.02} and the older drivers \meanSD{0.19}{0.03} at \pValue{.03}.
Similar to the distance to the lane center, the relative drift velocity statistics indicate higher values for older drivers, while no gender-specific effects are present. However, the performed one-way ANOVAs and Tukey Post-hoc test showed only significant mean differences between the younger and older drivers. The means of the Absmax values differ significantly between the groups \anova{Fisher's}{2}{59}{5.62}{0.01} while younger drivers \meanSD{1.50}{0.15} showed lower values compared to the older drivers \meanSD{1.74}{0.19} at \pValue{.01}. The age groups also affect the mean SD values significantly \anova{Fisher's}{2}{59}{5.21}{0.01}. Younger subjects \meanSD{0.70}{0.07} thereby had lower values than the older ones \meanSD{0.82}{0.10} at \pValue{.01}. Similarly, the ANOVAs report significantly higher values for the oldest group for the RMS \anova{Fisher's}{2}{59}{5.92}{0.01}, IDR \anova{Fisher's}{2}{59}{4.39}{0.02} and Median \anova{Fisher's}{2}{59}{3.67}{0.03} indicators.

\subsubsection*{G-G Envelope}
To form the Mean, Max, 95$^{th}$, and 75$^{th}$ percentile envelopes of the measured G-G diagrams, in the following, the angle tolerance $\delta_r$ is set to \SI{10}{\degree} and the rotation stride $\delta_s$ to \SI{15}{\degree} which introduces an angular overlap of \SI{5}{\degree}. In total, each envelope consists of $N_E=24$ reference points.
The resulting set of variables is sufficiently related for factor analysis, as indicated by Bartlett's Test of Sphericity \ChiSquarePL{276}{1604}.
In addition, the Kaiser-Meyer-Olkin measure reports a high overall strength of the relationships among variables \kmo{.81}.
PCA with Varimax rotation extracts two principal components based on the parallel analysis criterion. Component $\mathrm{PC}_1$ explains \SI{36.4}{\percent} and $\mathrm{PC}_2$ \SI{24.9}{\percent} of the variance, summing up to \SI{62.3}{\percent} in total.

To analyze the impact of each specific angle proportion on both principal components, their loadings are visualized in \autoref{fig:ggPCALoadings} using a polar G-G diagram. The visualization emphasizes that the two components can be attributed to a logical meaning. From this, it can be seen that $\mathrm{PC}_1$ focuses mainly on areas with positive longitudinal accelerations. In contrast, $\mathrm{PC}_2$ assigns a higher weighting to areas during braking and turning.
In \autoref{tab:ggEnvelopeKpiResults} the subject statistics of the four envelopes are summarized.
It is noticeable that $\mathrm{PC}_1$ shows higher standard deviations than $\mathrm{PC}_2$.
Regarding gender-specific differences, only the means of $\mathrm{PC}_1$ vary significantly.
A Mann-Whitney test revealed that male drivers \meanSD{0.597}{3.78} had significantly higher mean values of the first component of the mean envelope than female drivers \meanSD{-1.58}{2.24}, \mannWhitney{235}{.02}. The same pattern can be found for the max envelope. Male subjects \meanSD{0.50}{3.42} showed significantly higher mean values on $\mathrm{PC}_1$ than their female counterparts \meanSD{-1.33}{2.32}, \mannWhitney{241}{.03}.
According to a Yuen-Welch's test, also the $\mathrm{PC}_1$'s mean values of the 95$^{th}$ percentile envelope are gender-dependent, \yuenT{29.5}{3.25}{0.003}. The mean values for men \meanSD{0.59}{3.38} was significantly higher than for women \meanSD{-1.55}{1.56}.
Apart from this, neither gender nor age-group-specific dependencies could be found.

%
%

\begingroup

\setlength{\tabcolsep}{3pt} 

\begin{table}[]
    \caption{Statistics of the PCA componetns of the GG-Envelope}
    \centering
    \scriptsize
\begin{tabular}{llcccc}
\toprule
\textbf{Envelope} & \textbf{Component} & \textbf{Median} & \textbf{SD} & \textbf{Min} & \textbf{Max} \\ \midrule
Mean              & $\mathrm{PC}_1$                      & -0.25           & 3.55        & -7.31        & 12.21        \\
                  & $\mathrm{PC}_2$                      & -0.05           & 1.61        & -3.22        & 4.78         \\
Max               & $\mathrm{PC}_1$                      & -0.46           & 3.25        & -5.01        & 9.56         \\
                  & $\mathrm{PC}_2$                      & -0.15           & 1.76        & -4.01        & 4.25         \\
75$^{th}$ percentile  & $\mathrm{PC}_1$                      & -0.41           & 3.22        & -6.40        & 11.76        \\
                  & $\mathrm{PC}_2$                      & 0.09            & 1.50        & -4.26        & 4.41         \\
95$^{th}$ percentile  & $\mathrm{PC}_1$                      & -0.33           & 3.13        & -4.66        & 13.22        \\
                  & $\mathrm{PC}_2$                      & -0.17           & 1.64        & -2.72        & 4.49         \\ \bottomrule
\multicolumn{6}{l}{\footnotesize{Mean values are all zero. }} \\
\end{tabular}
\label{tab:ggEnvelopeKpiResults}
\end{table}

\endgroup

\subsubsection*{Stationary Cornering Behavior}
The statistic results of the curve cutting gradient, derived from quasi-stationary cornering, are listed in \autoref{tab:ccgKpiResults}. Noteworthy is that men, in contrast to some women, show almost no negative curve cutting. In other words, they do not tend to drift to the outside of the curve.
Similarly, there are differences in global offset values between the sexes, but these are not statistically significant.
Furthermore, no dependencies were found with regard to age groups.
This suggests that curve-cutting behavior depends individually on the driver and is not linked to socio-demographic factors.
\begingroup

\setlength{\tabcolsep}{3pt} 

\begin{table}[]
    \caption{Results of the Curve Cutting Gradient indicators}
    \centering
    \scriptsize
\begin{tabular}{llccccc}
\toprule
\textbf{KPI} & \textbf{Gender} & \textbf{Mean}  & \textbf{SD} & \textbf{Min} & \textbf{Max} \\
\midrule
Curve Cutting Gradient $\mathrm{CCG}$          & female          & 0.079          & 0.063       & -0.089           & 0.161            \\
             & male            & 0.099          & 0.049       & -0.002           & 0.199            \\
Global Offset $\mathrm{CCG}_0$       & female          & -0.009         & 0.107       & -0.170           & 0.213            \\
             & male            & 0.039          & 0.107       & -0.239           & 0.258            \\
95\% Confidence Interval $\mathrm{CI}_{d_{\mathrm{CL}}}$       & female          & 0.703          & 0.197       & 0.417            & 1.179            \\
             & male            & 0.687          & 0.140       & 0.400            & 1.067            \\
\bottomrule
\end{tabular}
\label{tab:ccgKpiResults}
\end{table}

\endgroup

\subsubsection*{Transient Cornering Behavior}
\autoref{tab:TRKpiResults} contains the statistical characteristics of the trajectory classes and their intensity values. 
Biased inner and center are the most prominent classes.
This distribution is consistent with the results of \cite{barendswaard2019classification}.
Across all classes, the minimum percentage values are zero except for the center (\SI{5}{\percent}). This means that each subject did not leave the center tolerance band in at least one curve and no curve cutting occurred.
In addition to the negative $\mathrm{CCG}$ values of the stationary cornering behavior, the existence of the counter-cutting class of up to \SI{12}{\percent} verifies that some subjects tend to drift to the outside of the curve. Compared to the cutting class, however, the intensity values are considerably lower.
After analyzing \num{16} class percentages and intensity values per subject, independent-sample t-tests revealed significant differences between the sexes for only two features.
For the Early Cutting class percentage, men \meanSD{4.99}{4.92} showed significantly higher mean values than women \meanSD{2.02}{2.53}, \yuenT{30.2}{2.11}{.043}. Also, the mean intensity values \meanSD{21.20}{30.8} of the Early Cutting class were significantly higher than for the female drivers \meanSD{7.06}{12.7}, \yuenT{35.6}{2.385}{.023}.
Similarly, a Mann-Whitney test (\mannWhitney{238}{.020}) proved that male drivers \meanSD{5.51}{4.45} had significantly higher mean percentages for the Early Counter class than female subjects \meanSD{2.97}{5.02}.
Again, according to a Mann-Whitney test, men \meanSD{33.53}{93.0} demonstrated significantly higher mean intensities for early counter compared to women \meanSD{9.09}{20.7}, \mannWhitney{219}{.008}.


\subsection{Correlation}
One objective of this study is to examine the correlations between the assigned driving styles from the self-assessment and the objective measures collected.
Therefore, \num{1080} Pearson correlations between the six MDSI driving scores and the \num{180} driving behavior indicators including all statistics of $a_x$, $a_y$, $d_{\mathrm{CL}}$, $v_{\mathrm{drift}}$, $k_x$, $k_y$, both components of the G-G Envelope, all features of the $\mathrm{CCG}$, trajectory classes, and trajectory intensities are calculated.
To verify that the effects can be attributed entirely to the driving style scores, partial correlations were computed controlling for age and gender similar to \cite{van2018relation,holman2015romanian}.
For rural roads, only \num{31} features show highly significant ($p < .01$) correlations and are summarized in \autoref{tab:correlationResults}. Overall, medium effect sizes occur.
There was a significant negative relationship between the Angry score and the 95$^{th}$ percentile of the G-G Envelope's $\mathrm{PC}_2$. Considering the shape of factor loadings in \autoref{fig:ggPCALoadings} there is a high probability that angry drivers reduce lateral accelerations during braking.
The score of the Risky driving style shows significant correlations with the statistics of longitudinal accelerations. The overall positive effect sizes indicate that drivers who rank high on this score are more tolerant towards high acceleration values and their dispersion. This is in line with the positive, significant correlations of the risky score and the G-G Envelope's $\mathrm{PC}_1$. In this case, higher mean and 95$^{th}$ percentile values of lateral and positive longitudinal accelerations are likely.

Most of the significant correlations were found for the Anxious score. For the lateral acceleration and longitudinal jerk statistics, medium negative relations exist with the anxiety measures. Contrary to this, there is a positive correlation between the Anxious score and the relative drift velocity and intensity values of the oscillating cutting class. Drivers who show high values in this score reduce the occurring lateral accelerations and decrease acceleration changes on the one hand. On the other hand, they tend to drift away from the lane center with higher velocities and oscillate more during cutting.
With the Dissociative score, the relative drift velocity, IDR values of the distance to the lane center, and the percentage of the early counter class are significantly correlated. In addition to faster drifting from the lane center, the measured spread values of $d_{\mathrm{CL}}$ increase with higher factor scores. 
These rather negative driving behaviors can be attributed to the tendency to be easily distracted while driving which leads to driving errors \cite{taubmann2004}.
For the Careful score, only the minimum lateral acceleration is significantly correlated. This inverse relationship means that careful drivers allow higher acceleration towards the right border of the lane and thus away from the oncoming lane. 
Finally, the Distress Reduction factor score was positively correlated with the occurrence of the late counter trajectory class.

Overall, it is evident that acceleration and jerk values account for the majority of the correlations.
This is in line with previous works, where these driving indicators are strongly linked with the human driving style \cite{van2018relation, murphey2009driver, rath2019lane, feng2018driving,moosavi2021driving,bejani2019convolutional, lv2018driving, kovaceva2020identification}. 
Beyond these more high-level KPIs, metrics specific to lateral driving behavior, such as the relative drift velocity, distance to lane center, and stationary and transient cornering behavior are almost unrepresented.
This can be explained by the high evidence, that humans are usually incapable of assessing their objective driving behavior and often overestimate their performance \cite{amado2014accurately, mynttinen2009novice, delhomme1991comparing,goszczynska1989self,sivak1989cross, svenson1981we, freund2005self, victoir2005learning}.

\begingroup

\setlength{\tabcolsep}{3pt} 

\begin{table}[t]
\caption{Summary of the highly significant ($p < .01$) partial correlations between the six driving styles and the objective driving behavior indicators.}
\centering
\scriptsize
\begin{tabular}{llllc}
\toprule
\textbf{Factor Score}    & \textbf{KPI}                & \textbf{Statistic}  & \textbf{Avg} & \textbf{Correlation} \\ 
\midrule
Angry              & GG Envelope                 &  $\mathrm{PC}_2$                 & 95$^{th}$ percentile & -0.334$^{**}$             \\
                   &                             &                     &              &                      \\
Risky              & Longitudinal Acceleration   & Absmax              & Median       & 0.375$^{**}$              \\
                   &                             & SD                  & Median       & 0.392$^{**}$              \\
                   &                             & RMS                 & Mean         & 0.350$^{**}$              \\
                   &                             & RMS                 & Median       & 0.410$^{**}$              \\
                   &                             & IDR                 & Median       & 0.469$^{***}$             \\
                   & GG Envelope                 & $\mathrm{PC}_1$                 & Mean         & 0.350$^{**}$              \\
                   &                             & $\mathrm{PC}_1$                 & 95$^{th}$ percentile          & 0.370$^{**}$              \\
                   &                             &                     &              &                      \\
Anxious            & Relative Drift Velocity     & Max                 & Mean         & 0.338$^{**}$              \\
                   & Lateral Acceleration        & Absmax              & Mean         & -0.392$^{**}$             \\
                   &                             & Absmax              & Median       & -0.336$^{**}$             \\
                   &                             & Max                 & Mean         & -0.411$^{**}$             \\
                   &                             & Max                 & Median       & -0.344$^{**}$             \\
                   &                             & Mean                & Mean         & -0.371$^{**}$             \\
                   &                             & Rms                 & Mean         & -0.376$^{**}$             \\
                   &                             & IDR                 & Median       & -0.338$^{**}$             \\
                   &                             & Median              & Mean         & -0.360$^{**}$             \\
                   & Longitudinal Jerk           & SD                  & Median       & -0.368$^{**}$             \\
                   &                             & RMS                 & Median       & -0.368$^{**}$             \\
                   &                             & IDR                 & Mean         & -0.374$^{**}$             \\
                   &                             & IDR                 & Median       & -0.385$^{**}$             \\
                   & Trajectory Intensity        & oscillating cutting & Mean         & 0.404$^{**}$              \\
                   &                             & oscillating cutting & Median       & 0.404$^{**}$              \\
                   &                             &                     &              &                      \\
Dissociative       & Relative Drift Velocity     & Absmax              & Median       & 0.393$^{**}$              \\
                   &                             & Max                 & Mean         & 0.388$^{**}$              \\
                   &                             & IDR                 & Median       & 0.333$^{**}$              \\
                   & Trajectory Class Percentage & early counter       &              & 0.350$^{**}$              \\
                   & Distance to Lane Center      & IDR                 & Median       & 0.334$^{**}$              \\
                   &                             &                     &              &                      \\
Careful            & Lateral Acceleration        & Min                 & Mean         & -0.363$^{**}$             \\
                   &                             & Min                 & Median       & -0.342$^{**}$             \\
                   &                             &                     &              &                      \\
Distress Reduction & Trajectory Class Percentage & late counter        &              & 0.422$^{***}$             \\ 
\bottomrule
\multicolumn{5}{l}{\footnotesize{Partial Pearson correlations, two-tailed test, controlling for age and gender}} \\
\multicolumn{5}{l}{\footnotesize{$^{**}~p < .01$, $^{***}~p < .001$}} \\
\end{tabular}
\label{tab:correlationResults}
\end{table}

\endgroup

\section{Discussion}

This study aimed to investigate the differences in human lateral driving behavior on rural roads to derive fine-grained descriptive indicators.
The more general, statistic-based indicators, like the max values of lateral jerks, used in previous works are not directly integrable for advanced driving functions that can be potentially adapted towards the specific human driver.
In addition, we wanted to evaluate if driving style information derived from self-reports correlates with the objective measurement data and can therefore be used for driving style adaption or the identification of different curve-negotiation types.

Regarding the dependencies of the behavior indicators and the subjects' socio-demographics, most relationships were found for the basic indicators.
Compared to the older age group, younger drivers showed significantly higher values for the mean longitudinal accelerations, standard deviation, and maximum of the lateral accelerations.
This is in line with previous studies, that indicate a negative correlation between age and driving speed \cite{boyce2002instrumented,farah2011age,starkey2016role,taubman2004multidimensional}.
The somewhat more leisurely driving style of older test subjects also leads to lower acceleration values.
The presumed more aggressive driving style of men was also partially demonstrated in this study, as they showed significantly higher mean jerk values in the lateral direction.
Interestingly, the maximum jerk values for accelerations and brakings were higher for female drivers.
One explanation for this might be the assumed tendency of women to be more anxious, dissociative, and distress-reduction drivers \cite{taubman2016value,poo2013study,taubman2004multidimensional,holland2010differential,poo2013reliability,wang2018effect,taubman2016multidimensional,long2019reliability,starkey2016role,gwyther2012effect}.
Anxious drivers react more impulsively to changing driving situations, which is reflected, for example, in stronger braking.

The results revealed a general curve-driving tendency of the subjects to drive with a lateral position slightly shifted of \SI{0.15}{\metre} to the right side of the lane. An explanation for this is given in \cite{ding2014driver}: perceived risk sensations vary significantly between the right and left sides of the lane. Drivers are more sensitive to left-side departures, initiating early corrections while being more tolerant of right-side lane departures. Interestingly this mean shift was found to be higher for male drivers. Moreover, the maximum distance to the lane center during cornering showed significant age-group dependencies. Older drivers demonstrated the highest offsets compared to the young and middle-aged participants.
One reason for this could be the higher tolerance to higher lane center deviations that come with more driving experience.
In contrast, however, the increased values can also be explained by the reduced reaction capabilities of older drivers.
Also in \cite{ghasemzadeh2018utilizing} the drivers' age turned out to significantly impact the lane-keeping performance.
Two indications that point in this direction are the significantly lower standard deviations of the lane deviations of the younger driver group and the significantly higher mean and standard deviations of the relative drift speeds of the older driver group.
Moreover, heightened lane position variability may suggest distracted driving, associated with a dissociative driving style \cite{just2008decrease}.

Comparatively few age- and gender-specific dependencies were found for the lower-level driving indicators. The analysis of the G-G Envelope reversals, that male drivers showed significantly higher values for the mean, $95^{th}$ percentile, and maximum values for $\mathrm{PC}_1$, which focuses mainly on areas with positive longitudinal accelerations. Combined with the distribution of the two factor scores in \autoref{fig:ggPCALoadings} it can be concluded that the male subjects make more use of the dynamic reserves of the vehicle than the female subjects.
Regarding the quasi-stationary and transient curve-driving behavior it can be concluded that pure lane-center driving is uncommon. This is also consistent with the findings from \cite{gordon2014modeling,rossner2020care,barendswaard2019classification}.
Based on convenience, drivers tend to choose a shorter trajectory when navigating curves \cite{ding2014driver}. 
For these KPIs, only relationships between the sexes and the percentage of trajectory classes early cutting and early counter were found.
In both cases, male drivers exhibit significantly higher class occurrences.
That this tendency is not based on coincidence is proven by the significantly higher intensity values of these classes, which is more than three times higher compared to female drivers.
This means that drivers consciously choose this type of curve cutting and also distinctly perform it.

To evaluate the extent the lateral driving behavior indicators can be predicted using driving styles derived from self-reports, we performed a correlation analysis between the calculated factor scores of the MDSI and the proposed indicators.
The outcomes of the present study align with existing literature, revealing consistent correlations between individuals' self-reported driving style scores and their actual driving behavior.
This affirms that the results derived from the Multidimensional Driving Style Inventory hold predictive value for the actual driving behavior observed in real-world scenarios, which was also concluded in \cite{kaye2018comparison,van2018relation,taubman2016value, zhao2012investigation,lajunen2003can,rengifo2021driving}.
This is interesting in that the questionnaire does not include explicit questions about driving behavior per se, but assesses driving tendencies and perceptions shaped by the driver's driving experience across various scenarios.
These aspects are intricately linked to the driver's personality traits \cite{taubman2004multidimensional} and the underlying driving styles exhibit relative stability over time \cite{saad2004behavioural,sagberg2015review}.
The majority of the correlations found, however, predominantly relate to the acceleration and jerk values.
This was to be expected inasmuch as these are one of the most used quantities for driving style analysis and classification \cite{martinez2017driving,murphey2009driver,chu2017curve,bellem2018comfort,bellem2016objective,bellem2017can,vilaca2017systematic,kanarachos2018smartphones,freyer2007ein}.
Drivers with distinct driving styles exhibit significant differences in lateral accelerations, with the maximum lateral acceleration depending on the driver's acceptable risk level and experience \cite{deng2020probabilistic}.
In this context, it is also plausible that the careful and the anxious factor score in particular shows significant negative correlations with the lateral acceleration and longitudual jerk statistics, while the risky score significantly positively correlates with longitudinal accelerations.
Drivers who scored high on the anxious or the dissociative score also showed higher relative drift velocities.
In other words, these drivers allow the vehicle to depart from the center of the lane more quickly, which is partly due to inattentiveness and partly due to slower reaction times.
This effect can also be observed in the positive correlation between the dissociative score and the interdecile range values of the distance to lane-center values and the significantly higher oscillating intensity values. Moreover, the dissociative factor score correlates positively with the early counter trajectory class-occurrence.
In this case, the vehicle begins the cornering maneuver at the outer edge of the curve but is corrected toward the center of the curve later on.
An opposite effect can be seen with careful drivers, which show a significant positive correlation towards the late counter class.
Trajectories of this class are characterized by the condition that the vehicle stays in the center of the lane for at least the first half of the curve.
While for the G-G Envelope, an expected positive correlation with the risk factor was found which is also consistent with the positive acceleration correlations, for the angry score surprisingly a significantly negative correlation with the second component of the PCA was revealed.
Considering the loading factors in \autoref{fig:ggPCALoadings}, it can be concluded that the more risky drivers tend to demonstrate less lateral acceleration during a braking maneuver, favoring a G-G diagram more shaped like a leaf compared to the other patterns defined in \cite{will2020methodological}.

In general, it can be observed that the majority of the group and self-rated driving style affiliations are found for the rather basic behavior indicators.
The absence of affiliations for in-depth lateral indicators can be attributed to the heterogeneous nature of driving styles among drivers.
There is substantial evidence, that every driver has its signature driving style \cite{dong2016characterizing, woo2018dynamic,brambilla2017comparison,sun2020intention,lin2014overview,tement2022assessment,kim2021driving}.
This highlights also the need to detect and predict the driving style to adapt to the specific driver's needs \cite{wang2010parallel, li2011cognitive,zhang2011data}, ultimately leading to a transformation from "driver adapts to car" to "car adapts to driver" \cite{zong2013neural}.
\section{Limitations}
A potential limitation of our study is that drivers were observed in an unfamiliar vehicle.
While over three-quarters of the participants reported prior experience driving an SUV, it's important to acknowledge the potential influence of the sports car used in this study.
However, in \cite{tanvir2018effect} the decomposition into inter-driver heterogeneity, inter-vehicle heterogeneity, and intra-driver-vehicle variability was investigated, concluding that the choice of vehicle does not significantly affect the driver's natural driving style.
We also attempted to mitigate this effect by allowing the subjects to experience the vehicle without specific instructions during the familiarization phase.
Moreover, due to the real-world conditions of this study on German roads, factors such as traffic density and the behavior of other vehicles couldn't be entirely controlled although no drives were carried out during high traffic volumes in the morning and evening.
Additionally, the presence of the co-pilot poses a limitation to the study.
One could argue that participants, upon realizing they are being observed during on-road driving, may consciously choose a more cautious driving style.
Despite our efforts, such as explicitly instructing drivers that the on-road session was not an exam and encouraging them to drive as usual, there remains the possibility of observer bias influencing the results.
Nevertheless, in certain studies \cite{quimby1999drivers,grayson2003risk}, the influence of this effect is perceived to be less significant than initially presumed.
In the current research, the investigation into self-reported driving styles was conducted with German drivers.
Although the MDSI has been applied in diverse countries, the factorial structure may exhibit slight variations.
The sample size in this study is considered insufficient for a robust factor analysis of the MDSI items.
For this reason, the factor divisions and loadings are taken from \cite{van2015measuring}.
However, due to country-specific differences between subjects in the two studies, these attributes may not be readily generalizable.
This can also be seen in the only mediocre fit of a Confirmatory Factor Analysis on our sample with $\chi^2 = 860$, $\mathrm{df} = 613$, $p < 0.001$, $\mathrm{CFI} = 0.688$, $\mathrm{TLI} = 0.661$, $\mathrm{SRMR} = 0.110$, $\mathrm{RMSEA} = 0.0807$, and $\mathrm{AIC} = 6059$.
When deriving conclusions from studies with relatively small sample sizes in combination with an unequal distribution of gender, such as the one reported here, caution is necessary.
Sampling errors can result in overestimates of the size of any true observed effects.
Moreover, the constrained statistical power may pose challenges in initially detecting such effects \cite{button2013power}.
However, it is worth noting that the internal consistency values of this study align well with those reported in studies with larger sample sizes.
\section{Conclusion}

In conclusion, we conducted a controlled driving study using a research vehicle equipped with a comprehensive multimodal sensor set to analyze the lateral driving behavior of human drivers, specifically emphasizing rural roads.
To objectify the driving behavior we proposed novel indicators to evaluate stationary and transient curve-negotiation that can be directly applied within the development process of lateral driving functions. 
To evaluate if these indicators can be predicted using self-reports of driving styles, we proposed the MDSI-DE, the German version of the Multidimensional Driving Style Inventory enriched with additional items.
A correlation analysis between the calculated factor scores of the MDSI and the proposed indicators revealed modest but significant correlations primarily with the basic, statistic-based, features derived from the lateral and longitudinal acceleration and jerk values.
However, these driving style measures did not prove to serve as proxies for the more in-depth lateral driving behavior, like curve cutting or trajectory preference.
The lack of associations with detailed lateral indicators can be attributed to the diverse nature of human driving styles.

We recommend that future research focus on the incorporation of more detailed driving behavior indicators beyond the pure acceleration and jerk values into the personalization process of lateral driving functions.
The results indicate a high driver-heterogeneous regarding the curve-negotiation behavior.
An adaption towards the specific human driver could possibly mitigate existing acceptance barriers toward partially or fully autonomous driving and can improve the overall comfort experience.
To facilitate further research, we made the dataset including the anonymized socio-demographics and questionnaire responses, the raw vehicle measurements including labels, and the derived driving behavior indicators publicly available.

\newpage
\section*{LIST OF ABBREVIATIONS}

\begin{acronym}\itemsep-10pt
	\acro{ads}[ADS]{Aggressive Driving Scale}
	\acro{aic}[AIC]{Akaike Information Criteria}
	\acro{anova}[ANOVA]{analysis of variance}
	\acro{aq}[AQ]{Aggression Questionnaire}
	\acro{arca}[ARCA]{Automated Ride Comfort Assessment}
	\acro{avs}[AVs]{Autonomous Vehicles} 
	\acro{cas}[CAS]{Coronavirus Anxiety Scale}
	\acro{ccg}[CCG]{curve cutting gradient}
	\acro{csai}[CSAI-2]{Competitive State Anxiety Inventory 2}
	\acro{das}[DAS]{Driving Anger Scale}
	\acro{dbi}[DBI]{Driving Behaviour Inventory}
	\acro{dbq}[DBQ]{Driver Behavior Questionnaire}
	\acro{deq}[DEQ]{Driver Evaluation Questionnaire}
	\acro{dscf}[DSCF]{Dwass-Steel-Critchlow-Fligner} 
	\acro{dseq}[DSEQ]{Driver Self-Evaluation Questionnaire}
	\acro{dsq}[DSQ]{Driving Style Questionnaire}
	\acro{dvq}[DVQ]{Driving Vengeance Questionnaire}
	\acro{fms}[FMS]{Fast Motion Sickness Scale}
	\acro{idr}[IDR]{interdecile range}
	\acro{kpi}[KPI]{Key-Performance-Indicator}
	\acro{ldw}[LDW]{Lane Departure Warning}
	\acro{lks}[LKS]{Lane Keeping Systems}
	\acro{loc}[LoC]{Locus of Control}
	\acro{mdsi}[MDSI]{Multidimensional Driving Style Inventory}
	\acro{pca}[PCA]{principal component analysis}
	\acro{pss}[PSS]{Perceived Stress Scale}
	\acro{rdhs}[RDHS]{Reckless Driving Habits Scale}
	\acro{rms}[RMS]{root mean square}
	\acro{sd}[SD]{standard deviation}
	\acro{ssq}[SSQ]{Simulator Sickness Questionnaire}
	\acro{tas}[TAS]{Thrill and Adventure Seeking}
	\acro{tia}[TiA]{Trust in Automated Systems}

\end{acronym}



\section*{CREDIT AUTHORSHIP CONTRIBUTION STATEMENT}
\textbf{Johann Haselberger:} Conceptualization, Methodology, Software, Validation, Formal analysis, Investigation, Writing - original draft, and Writing - Review \& Editing. \newline
\textbf{Bernhard Schick:} Conceptualization and Writing - Review \& Editing. \newline
\textbf{Steffen Müller:} Conceptualization and Writing - Review \& Editing. 
\section*{ACKNOWLEDGMENTS}
The authors want to thank Porsche Engineering Group GmbH for the support and
for providing the test vehicle as part of a larger research project.
Additionally, the authors want to thank Fabian Diet for his
contributions during the subject management and test execution. 
\section*{DATA AVAILABILITY}
The dataset including the anonymized socio-demographics and questionnaire responses, the raw vehicle measurements including labels, and the derived driving style indicators is publicly available.
\section*{CONFLICTS OF INTEREST}
The authors declare that they have no known competing financial interests or personal relationships that could have appeared to influence the work reported in this paper. 

\newpage
\bibliographystyle{apa}

\bibliography{references}  

\newpage
\appendix
\section{Appendix}


\begin{figure}[h]
		\centering
		\small
		\begin{tabular}{cc}
			\includegraphics[width=.4\linewidth,valign=m,trim={0 0 -1cm 0}]{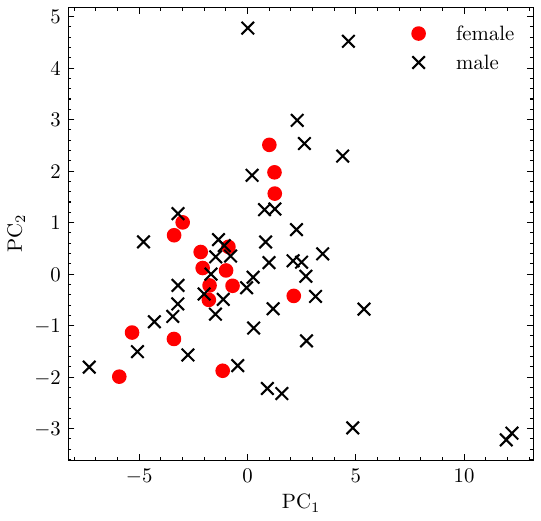} & 
			\includegraphics[width=.4\linewidth,valign=m]{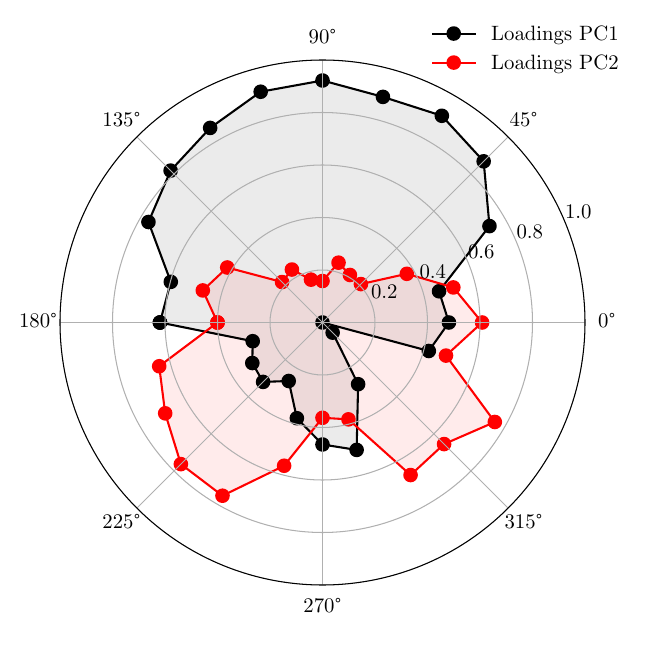} \\
			a) & b)  \\
		\end{tabular}
		\caption{a) Scatterplot of the $\mathrm{PC}_1$ and $\mathrm{PC}_2$ values of the mean envelope for female and male drivers on rural roads. b) Loadings of the PCA for the Mean-Envelope visualized in the GG-Diagram. PC1 focuses more on areas with positive longitudinal accelerations, while PC2 assigns a higher weighting to areas during braking and turning.}
\label{fig:ggPCALoadings}
\end{figure}


\begingroup

\setlength{\tabcolsep}{3pt} 

\begin{table}[h]
    \caption{\small{Self reported car usage and annual mileage split into Women(N=17) and Men (N=45).}}
    \centering
    \scriptsize
    \begin{tabular}{lllllll}
    \toprule
    Car usage            &        &    &  & Annual mileage       &        &    \\ \cline{1-3} \cline{5-7} 
    Response             & Gender & N  &  & Response           & Gender & N  \\ \midrule
    Daily                & female & 9  &  & 0 - 5k           & female & 4  \\
                         & male   & 21 &  &                    & male   & 3  \\
    Several times a week & female & 7  &  & 5k - 10k       & female & 4  \\
                         & male   & 16 &  &                    & male   & 9  \\
    Weekly               & female & 0  &  & 10k -20k       & female & 9  \\
                         & male   & 5  &  &                    & male   & 19 \\
    Few times a month    & female & 1  &  & 20k - 30k      & female & 0  \\
                         & male   & 3  &  &                    & male   & 12 \\
                         &        &    &  & \textgreater 30k & female & 0  \\
                         &        &    &  &                    & male   & 2  \\ \bottomrule
    \end{tabular}
    \label{tab:participants02}
    \end{table}

    \endgroup

\afterpage{
\onecolumn
\footnotesize
\renewcommand{\doublerulesep}{0pt}
\begin{xltabular}{1\linewidth}{l >{\arraybackslash}X ll}
\caption{
  Summary of all MDSI-DE questions divided into the eight factors including statistics and reliability. The factor assignments are initially inherited from \cite{van2015measuring} and gradually enriched with remaining items from \cite{taubmann2004}, \cite{poo2013reliability}, and self-defined items. Participants scored their agreement on each item using a 6-point Likert scale, ranging from 1 (not at all) to 6 (very much).
  } \\
\toprule
\multicolumn{2}{l}{\textbf{MDSI items}} & \textbf{M} & \textbf{SD} \\
\midrule
\endfirsthead
\toprule
\multicolumn{2}{l}{\textbf{MDSI items}} & \textbf{M} & \textbf{SD} \\ 
\midrule
\endhead
\midrule
\multicolumn{4}{c}{\textit{Continued on next page\ldots}}\\
\endfoot
\endlastfoot
&  &  &  \\
\multicolumn{2}{l}{\textbf{Factor 1 - Angry Driving (Cronbach's alpha 0.721)}} & \textbf{} & \textbf{} \\
43 & Honk my horn at others$^a$ & 2.16 & 0.995 \\
\textit{} & \textit{Ich hupe andere Fahrer an} & \textit{} & \textit{} \\
3 & Blow my horn or "flash" the car in front as a way of expressing frustrations$^a$ & 1.90 & 1.067 \\
\textit{} & \textit{Ich benutze die Hupe oder die Lichthupe, um meine Frustration   gegenüber dem vorausfahrenden Auto auszudrücken} & \textit{} & \textit{} \\
28 & When someone does something on the road that annoys me, I   flash them with high beam$^a$ & 1.89 & 1.026 \\
\textit{} & \textit{Wenn jemand auf der Straße etwas macht, dass mich verärgert,   blende ich ihn mit dem Fernlicht} & \textit{} & \textit{} \\
12 & Swear at other drivers$^a$ & 3.52 & 1.225 \\
\textit{} & \textit{Ich schimpfe über andere Fahrer} & \textit{} & \textit{} \\
17 & When a traffic light turns green and the car in front of me doesn’t get going immediately, I try to urge the driver to move on$^a$ & 2.60 & 1.420 \\
\textit{} & \textit{Wenn eine Ampel grün wird und der Fahrer vor mir nicht sofort   losfährt, versuche ich den Fahrer zum losfahren zu bewegen} & \textit{} & \textit{} \\
13 & When a traffic light turns green and the car in front of me   doesn’t get going, I wait for a while until it moves$^a$ {[}-{]} & 4.47 & 1.399 \\
\textit{} & \textit{Wenn eine Ampel grün wird und das Auto vor mir nicht losfährt,   warte ich einfach eine Weile, bis es sich bewegt} & \textit{} & \textit{} \\
2 & Purposely tailgate other drivers$^a$ & 1.92 & 1.029 \\
\textit{} & \textit{Ich fahre absichtlich zu dicht auf} & \textit{} & \textit{} \\
45 & Arguing with other drivers or pedestrians$^c$ & 1.56 & 0.898 \\
\textit{} & \textit{Ich streite mich mit anderen Fahrern oder Fußgängern} & \textit{} & \textit{} \\
46 & Get angry with people driving slow in the fast lane$^c$ & 4.40 & 1.299 \\
\textit{} & \textit{Ich ärgere mich über Leute, die auf der linken Spur langsam   fahren} & \textit{} & \textit{} \\
 &  &  &  \\
\multicolumn{2}{l}{\textbf{Factor 2 - Risky Driving (Cronbach's alpha 0.872)}} & \textbf{} & \textbf{} \\
44 & Enjoy the excitement of dangerous driving$^a$ & 1.97 & 1.267 \\
\textit{} & \textit{Ich genieße die Aufregung eines gefährlichen Fahrstils} & \textit{} & \textit{} \\
22 & Like to take risks while driving$^a$ & 1.89 & 0.925 \\
\textit{} & \textit{Ich gehe gerne Risiken während der Fahrt ein} & \textit{} & \textit{} \\
24 & Like the thrill of flirting with death or disaster$^a$ & 1.60 & 1.016 \\
\textit{} & \textit{Ich brauche den Nervenkitzel mit der Gefahr zu spielen} & \textit{} & \textit{} \\
6 & Enjoy the sensation of driving on the limit$^a$ & 2.44 & 1.522 \\
\textit{} & \textit{Ich genieße das Gefühl, am Limit zu fahren} & \textit{} & \textit{} \\
29 & Get a thrill out of breaking the law$^a$ & 1.58 & 1.064 \\
 & \textit{Ich finde es aufregend, das Gesetz zu brechen} &  &  \\
47 & Enjoy the power of the engine$^c$ & 3.92 & 1.53 \\
 & \textit{Ich genieße die Kraft des Motors} &  &  \\
48 & Enjoy shifting gears quickly$^c$ & 3.27 & 1.48 \\
 & \textit{Ich genieße es, schnell die Gänge zu wechseln} &  &  \\
49 & Feel the car asking for more speed$^c$ & 2.29 & 1.40 \\
 & \textit{Ich fühle, dass das Fahrzeug förmlich nach mehr   Geschwindigkeit verlangt} &  &  \\
50 & Drive faster when a vehicle is trying to pass me$^c$ & 1.73 & 1.13 \\
 & \textit{Ich fahre schneller, wenn ein Fahrzeug versucht, mich zu   überholen} &  &  \\
51 & I enjoy the variation on curvy roads$^d$ & 4.37 & 1.50 \\
 & \textit{Ich genieße die Abwechslung auf kurvenreichen Straßen} &  &  \\
 &  &  &  \\
\multicolumn{2}{l}{\textbf{Factor 3 - Anxious Driving (Cronbach's alpha 0.678)}} & \textbf{} & \textbf{} \\
31 & Feel nervous while driving$^a$ & 1.65 & 0.977 \\
 & \textit{Ich fühle mich beim Fahren nervös} &  &  \\
10 & Driving makes me feel frustrated$^a$ & 1.63 & 1.012 \\
 & \textit{Fahren frustriet mich} &  &  \\
40 & Feel comfortable while driving$^a$ {[}-{]} & 5.27 & 1.011 \\
 & \textit{Ich fühle mich beim Fahren wohl} &  &  \\
4 & Feel I have control over driving$^a$ {[}-{]} & 5.13 & 1.094 \\
 & \textit{Ich habe das Gefühl Kontrolle zu haben, wenn ich fahre} &  &  \\
25 & It worries me when driving in bad weather$^a$ & 2.18 & 1.274 \\
 & \textit{Fahren bei schlechtem Wetter beunruhigt mich} &  &  \\
33 & Feel distressed while driving$^a$ & 1.24 & 0.592 \\
 & \textit{Ich fühle mich beim Fahren bedrückt} &  &  \\
41 & Always ready to react to unexpected manoeuvres by other drivers$^a$ {[}-{]} & 5.11 & 0.630 \\
 & \textit{Ich bin stets in der Lage auf unerwartetes Fahrverhalten   anderer zu reagieren} &  &  \\
8 & While driving, I try to relax myself$^a$ {[}-{]} & 4.18 & 1.235 \\
 & \textit{Wenn ich fahre, versuche ich mich zu entspannen} &  &  \\
52 & Driving on narrow rural roads overwhelms me$^d$ & 2.05 & 1.40 \\
 & \textit{Fahren auf engen Landstraßen überfordert mich} &  &  \\
  &  &  &  \\
\multicolumn{2}{l}{\textbf{Factor 4 - Dissociative Driving (Cronbach's alpha 0.693)}} & \textbf{} & \textbf{} \\
35 & Attempt to drive away from traffic lights in third gear (or on   the neutral mode in automatic cars)$^a$ & 1.10 & 0.433 \\
 & \textit{Ich versuche, im dritten Gang (oder im Leerlauf bei   Automatikfahrzeugen) von der Ampel wegzufahren} &  &  \\
27 & Forget that my lights are on full beam until flashed by   another motorist$^a$ & 2.02 & 0.983 \\
 & \textit{Ich vergesse, dass mein Fernlicht an ist, bis andere   Autofahrer mich darauf hinweisen} &  &  \\
39 & Nearly hit something due to misjudging my gap in a parking   lot$^a$ & 2.02 & 1.032 \\
 & \textit{Ich hätte fast etwas angefahren, weil ich meinen Abstand auf   einem Parkplatz falsch eingeschätzt habe} &  &  \\
36 & Plan my route badly, so that I hit traffic that I could have   avoided$^a$ & 2.52 & 1.211 \\
 & \textit{Ich plane meine Route schlecht, so dass ich auf Verkehr stoße,   den ich hätte vermeiden können} &  &  \\
34 & Intend to switch on the windscreen wipers, but switch on the   lights instead$^a$ & 1.35 & 0.726 \\
 & \textit{Anstatt den Scheibenwischer anzumachen betätige ich aus   Versehen den Blinker} &  &  \\
20 & Fix my hair / makeup while driving$^a$ & 1.15 & 0.438 \\
 & \textit{Ich mache meine Haare oder schminke mich, wenn ich fahre} &  &  \\
21 & Distracted or preoccupied, and suddenly realize the vehicle   ahead has slowed down, and have to slam on the breaks to avoid a collision$^a$ & 2.21 & 0.943 \\
 & \textit{Um einen Zusammenstoß zu vermeiden muss ich stark bremsen,   weil ich abgelenkt war und nicht bemerkt habe, dass das Fahrzeug vor mir   bremst} &  &  \\
5 & Drive through traffic lights that have just turned red$^a$ & 1.58 & 0.860 \\
 & \textit{Ich überfahre Ampeln, die gerade auf rot geschaltet haben} &  &  \\
19 & When someone tries to skirt in front of me on the road, I   drive in an assertive way in order to prevent it$^a$ & 1.92 & 1.121 \\
 & \textit{Wenn jemand versucht auf der Fahrspur vor mir einzuscheren,   fahre ich selbstbewusst, um dies zu verhindern} &  &  \\
30 & Misjudge the speed of an oncoming vehicle when passing$^b$ & 1.85 & 0.973 \\
 & \textit{Ich schätze die Geschwindigkeit eines entgegenkommenden   Fahrzeugs beim Überholen falsch ein} &  &  \\
15 & Lost in thoughts or distracted, I fail to notice someone at   the pedestrian crossings$^b$ & 1.66 & 0.700 \\
 & \textit{In Gedanken versunken oder abgelenkt, übersehe ich jemanden an   einem Fußgängerüberweg} &  &  \\
 &  &  &  \\
\multicolumn{2}{l}{\textbf{Factor 5 - Careful Driving (Cronbach's alpha 0.729)}} & \textbf{} & \textbf{} \\
42 & Tend to drive cautiously$^a$ & 4.15 & 1.171 \\
 & \textit{Ich neige zu vorsichtigem Fahren} &  &  \\
14 & Drive cautiously$^a$ & 4.82 & 0.915 \\
 & \textit{Ich fahre vorsichtig} &  &  \\
23 & Base my behaviour on the motto "better safe than   sorry"$^a$ & 4.79 & 1.088 \\
 & \textit{Ich fahre nach dem Motto "Vorsicht ist besser als   Nachsicht"} &  &  \\
7 & On a clear freeway, I usually drive at or a little below the   speed limit$^a$ & 3.55 & 1.734 \\
 & \textit{Auf einer freien Autobahn fahre ich in der Regel innerhalb   oder knapp unterhalb des Tempolimits} &  &  \\
41 & Always ready to react to unexpected manoeuvres by other   drivers$^a$ & 5.11 & 0.630 \\
 & \textit{Ich bin stets in der Lage auf unerwartetes Fahrverhalten   anderer zu reagieren} &  &  \\
53 & Wait patiently when not having right of way$^c$ & 3.40 & 1.40 \\
 & \textit{Ich warte geduldig, wenn ich nicht Vorfahrt habe} &  &  \\
54 & Wait patiently when you cannot advance the traffic$^c$ & 4.95 & 1.19 \\
 & \textit{Ich warte geduldig, wenn ich den Verkehr nicht vorantreiben   kann} &  &  \\
 &  &  &  \\
\multicolumn{2}{l}{\textbf{Factor 6 - Distress - Reduction Driving Style (Cronbach's alpha 0.504)}} & \textbf{} & \textbf{} \\
1 & Do relaxing activities while driving$^a$ & 1.84 & 1.162 \\
 & \textit{Ich mache entspannende Aktivitäten während der Fahrt} &  &  \\
37 & Use muscle relaxation techniques while driving$^a$ & 1.44 & 0.861 \\
 & \textit{Ich nutze Muskelentspannungstechniken beim Autofahren} &  &  \\
26 & Meditate while driving$^a$ & 1.11 & 0.367 \\
 & \textit{Ich meditiere während der Fahrt} &  &  \\
11 & I daydream to pass the time while driving$^a$ & 2.68 & 1.491 \\
 & \textit{Ich habe Tagträume, um mir die Zeit während der Fahrt zu   vertreiben} &  &  \\
55 & Listen to music to relax while driving$^c$ & 4.94 & 1.42 \\
 & \textit{Ich höre Musik, um beim Fahren zu entspannen} &  &  \\
56 & Enjoy the landscape while driving$^c$ & 3.81 & 1.32 \\
 & \textit{Ich genieße die Landschaft beim Fahren} &  &  \\
57 & If time permits, I prefer to drive on the rural road instead   of the highway$^d$ & 3.26 & 1.70 \\
 & \textit{Wenn es die Zeit zulässt, fahre ich lieber über die   Landstraße, anstatt die Autobahn} &  &  \\
 &  &  &  \\
\multicolumn{2}{l}{\textbf{Factor 7 - High-Velocity Driving Style (Cronbach's alpha 0.630)}} & \textbf{} & \textbf{} \\
16 & In a traffic jam, I think about ways to get through the   traffic faster$^b$ & 2.90 & 1.58 \\
 & \textit{Im Stau denke ich darüber nach, wie ich schneller durch den   Verkehr komme} &  &  \\
9 & When in a traffic jam and the lane next to me starts to move,   I try to move into that lane as soon as possible$^b$ & 2.31 & 1.14 \\
 & \textit{Wenn ich im Stau stehe und die Spur neben mir beginnt sich zu   bewegen, versuche ich, so schnell wie möglich auf diese Spur zu wechseln} &  &  \\
17 & When a traffic light turns green and the car in front of me   doesn’t get going immediately, I try to urge the driver to move on$^a$ & 2.60 & 1.420 \\
 & \textit{Wenn eine Ampel grün wird und der Fahrer vor mir nicht sofort   losfährt, versuche ich den Fahrer zum losfahren zu bewegen} &  &  \\
2 & Purposely tailgate other drivers$^a$ & 1.92 & 1.029 \\
 & \textit{Ich fahre absichtlich zu dicht auf} &  &  \\
32 & Get impatient during rush hours$^b$ & 3.19 & 1.35 \\
 & \textit{Ich werde während der Rushhour ungeduldig} &  &  \\
5 & Drive through traffic lights that have just turned red$^a$ & 1.58 & 0.860 \\
 & \textit{Ich überfahre Ampeln, die gerade auf rot geschaltet haben} &  &  \\
 &  &  &  \\
\multicolumn{2}{l}{\textbf{Factor 8 - Patient Driving Style (Cronbach's alpha 0.330)}} & \textbf{} & \textbf{} \\
18 & At an intersection where I have to give right-of-way to   oncoming traffic, I wait patiently for cross-traffic to pass$^b$ & 5.37 & 0.945 \\
 & \textit{An einer Kreuzung, an der ich dem Gegenverkehr die Vorfahrt   gewähren muss, warte ich geduldig, bis der Querverkehr vorbeigefahren ist} &  &  \\
23 & Base my behaviour on the motto "better safe than   sorry"$^a$ & 4.79 & 1.088 \\
 & \textit{Ich fahre nach dem Motto "Vorsicht ist besser als   Nachsicht"} &  &  \\
13 & When a traffic light turns green and the car in front of me   doesn’t get going, I wait for a while until it moves$^a$ {[}-{]} & 4.47 & 1.399 \\
 & \textit{Wenn eine Ampel grün wird und das Auto vor mir nicht losfährt,   warte ich einfach eine Weile, bis es sich bewegt} &  &  \\
38 & Plan long journeys in advance$^b$ & 2.63 & 1.417 \\
 & \textit{Ich plane Fahrten lange im Voraus} &  &  \\
\bottomrule
\multicolumn{4}{l}{\footnotesize{$^a$ Items that belong to the validated version of the MDSI \cite{van2015measuring}}} \\
\multicolumn{4}{l}{\footnotesize{$^b$ Remaining items from the original version of the MDSI \cite{taubmann2004}}} \\
\multicolumn{4}{l}{\footnotesize{$^c$ Remaining items from the adapted MDSI version for Spanish drivers \cite{poo2013reliability}}} \\
\multicolumn{4}{l}{\footnotesize{$^d$ Own defined items}} \\
\multicolumn{4}{l}{{[}-{]} reversed item} \\
\label{tab:participants052}
\end{xltabular}
\twocolumn
}

\begingroup

\setlength{\tabcolsep}{3pt} 

\begin{table}[h]
    \caption{Mapping of the four-digit curve encoding to trajectory class labels, where \tCode{C} stands for center, \tCode{I} for inner, and \tCode{O} for outer curve driving.}
    \centering
    \scriptsize
    \begin{tabular}{ll}
\toprule
\textbf{Trajectory Class}      & \textbf{Curve Encodings}                                 \\
\midrule
Center$^a$              & \tCode{CCCC}                                   \\
Early Cutting$^b$       & \tCode{ICCC}, \tCode{IICC}                     \\
Early Counter$^b$       & \tCode{OCCC}, \tCode{OOCC}                     \\
Late Cutting$^b$        & \tCode{CCCI}, \tCode{CCII}                     \\
Late Counter$^b$        & \tCode{CCCO}, \tCode{CCOO}                     \\
Cutting$^a$             & \tCode{OOII}, \tCode{OIII}, \tCode{CICC}, \tCode{CCIC}, \tCode{CIIC}                   \\
Counter$^a$             & \tCode{IIOO}, \tCode{IOOO}, \tCode{COCC}, \tCode{CCOC}, \tCode{COOC}                   \\
Severe Cutting$^a$      & \tCode{OIIO}, \tCode{OIIC}, \tCode{OCIC}, \tCode{OCIO}, \tCode{OICO}, \tCode{OICC}, \tCode{OCII}, \tCode{COII} \\
Severe Counter$^a$      & \tCode{IOOI}, \tCode{IOOC}, \tCode{ICOC}, \tCode{ICOI}, \tCode{IOCI}, \tCode{IOCC}, \tCode{ICOO}, \tCode{CIOO} \\
Biased Inner$^a$        & \tCode{CIII}, \tCode{IIIC}, \tCode{IIII}                               \\
Biased Outer$^a$        & \tCode{COOO}, \tCode{OOOC}, \tCode{OOOO}                               \\
Oscillating$^b$         & \tCode{OCOI}, \tCode{ICIO}, \tCode{CIOC}, \tCode{COIC}, \tCode{CCIO}, \tCode{CCOI}             \\
Oscillating Cutting$^b$ & \tCode{CICI}, \tCode{ICIC}, \tCode{ICII}, \tCode{IICI}                         \\
Oscillating Counter$^b$ & \tCode{COCO}, \tCode{OCOC}, \tCode{OCOO}, \tCode{OOCO}                         \\
Slow Severe Cutting$^b$ & \tCode{OCCI}                                           \\
Slow Severe Counter$^b$ & \tCode{ICCO}                                          \\
\bottomrule
\multicolumn{2}{l}{\footnotesize{$^a$ Class adapted from \cite{barendswaard2019classification} }} \\
\multicolumn{2}{l}{\footnotesize{$^b$ New class derived from the measurements }} \\
\end{tabular}
\label{tab:trajectoryClassMapping}
\end{table}

\endgroup

\begingroup

\setlength{\tabcolsep}{3pt} 

\begin{table}[h]
\caption{Statistics of the non-Refined, refined factor scores, and the derived driving styles split by gender. The non-refined scores are calculated using Weighted Sum Scores while the refined scores are estimated by multiple regression \cite{distefano2009understanding}. N$_\subset$ denotes the number of associated subjects assigned to a driving style.}
 \centering
 \scriptsize
\begin{tabular}{llccccccccccc}
\toprule
\textbf{}                  & \textbf{}       & \multicolumn{5}{c}{\textbf{Non-Refined}}      &                                & \multicolumn{5}{c}{\textbf{Refined}}                                          \\ \cline{3-7} \cline{9-13} 
\textbf{Driving Style}     & \textbf{Gender} & \textbf{Mean} & \textbf{SD} & \textbf{Min} & \textbf{Max} & \textbf{N$_\subset$} & & \textbf{Mean} & \textbf{SD} & \textbf{Min} & \textbf{Max} & \textbf{N$_\subset$} \\
\midrule
Angry              & female          & 1.04          & 0.37        & 0.34         & 1.81         & 0 &                & -0.02         & 0.71        & -1.17        & 1.46         & 1                 \\
                           & male            & 1.03          & 0.50        & 0.24         & 2.63         & 1  &               & 0.01          & 0.98        & -1.26        & 3.17         & 6                 \\
Anxious            & female          & -0.53         & 0.43        & -1.13        & 0.21         & 0   &              & 0.70          & 1.32        & -0.70        & 3.76         & 7                 \\
                           & male            & -0.82         & 0.20        & -1.12        & -0.31        & 0   &              & -0.26         & 0.58        & -0.72        & 1.71         & 2                 \\
Careful            & female          & 2.53          & 0.36        & 1.58         & 3.02         & 16  &              & 0.04          & 0.70        & -1.97        & 1.09         & 2                 \\
                           & male            & 2.51          & 0.48        & 1.12         & 3.36         & 38 &               & -0.02         & 0.97        & -3.08        & 1.53         & 15                \\
Dissociative       & female          & 0.88          & 0.25        & 0.59         & 1.47         & 0  &               & 0.32          & 1.05        & -1.04        & 2.56         & 3                 \\
                           & male            & 0.76          & 0.20        & 0.46         & 1.24         & 0  &               & -0.12         & 0.76        & -1.27        & 1.75         & 4                 \\
Distress Reduction & female          & 1.07          & 0.35        & 0.62         & 1.77         & 0  &               & -0.11         & 0.99        & -0.72        & 2.71         & 3                 \\
                           & male            & 1.12          & 0.42        & 0.62         & 2.26         & 0   &              & 0.04          & 1.01        & -0.72        & 2.72         & 8                 \\
Risky              & female          & 1.16          & 0.57        & 0.73         & 2.78         & 1  &               & -0.32         & 0.69        & -0.84        & 1.70         & 1                 \\
                           & male            & 1.46          & 0.73        & 0.73         & 3.23         & 6   &              & 0.12          & 1.02        & -0.84        & 2.64         & 10                \\
\bottomrule
\end{tabular}
\label{tab:mdsiNonAndRefinedScores}
\end{table}

\endgroup

\begingroup

\setlength{\tabcolsep}{3pt} 

\begin{table}[]
    \caption{Statistics of the acceleration values. The results for the mean and median are omitted as these values converge to zero.}
    \centering
    \scriptsize
\begin{tabular}{llccccccccc}
\toprule
\textbf{}     & \textbf{}    & \multicolumn{4}{c}{\textbf{Longitudinal Acceleration} $a_x$}                 &  & \multicolumn{4}{c}{\textbf{Lateral Acceleration} $a_y$}                      \\ \cline{3-6} \cline{8-11} 
\textbf{Statistic} & \textbf{Avg} & \textbf{Mean} & \textbf{SD} & \textbf{Min} & \textbf{Max} &  & \textbf{Mean} & \textbf{SD} & \textbf{Min} & \textbf{Max} \\ 
\midrule
Absmax        & mean         & 1.221         & 0.176       & 0.917        & 1.970        &  & 2.536         & 0.193       & 2.091        & 2.997        \\
              & median       & 1.065         & 0.121       & 0.854        & 1.492        &  & 2.492         & 0.197       & 1.992        & 2.861        \\
Max           & mean         & 0.896         & 0.145       & 0.669        & 1.509        &  & 2.532         & 0.195       & 2.091        & 2.997        \\
              & median       & 0.824         & 0.109       & 0.593        & 1.308        &  & 2.490         & 0.196       & 1.992        & 2.861        \\
Min           & mean         & -0.884        & 0.102       & -1.289       & -0.744       &  & -0.241        & 0.137       & -0.625       & 0.072        \\
              & median       & -0.745        & 0.080       & -0.986       & -0.594       &  & -0.200        & 0.125       & -0.570       & 0.130        \\
SD            & mean         & 0.331         & 0.063       & 0.232        & 0.648        &  & 0.492         & 0.036       & 0.423        & 0.577        \\
              & median       & 0.285         & 0.038       & 0.216        & 0.418        &  & 0.472         & 0.035       & 0.402        & 0.533        \\
RMS           & mean         & 0.460         & 0.079       & 0.338        & 0.804        &  & 1.324         & 0.132       & 1.016        & 1.604        \\
              & median       & 0.370         & 0.059       & 0.276        & 0.584        &  & 1.291         & 0.137       & 0.998        & 1.574        \\
IDR           & mean         & 0.442         & 0.091       & 0.314        & 0.894        &  & 0.663         & 0.054       & 0.567        & 0.812        \\
              & median       & 0.362         & 0.049       & 0.277        & 0.536        &  & 0.615         & 0.048       & 0.518        & 0.730        \\
\bottomrule
\end{tabular}
\label{tab:accelKpiResults}
\end{table}

\endgroup

\begingroup

\setlength{\tabcolsep}{3pt} 

\begin{table}[]
    \caption{Statistics of the jerk values. The results for the mean and median are omitted as these values converge to zero.}
    \centering
    \scriptsize
\begin{tabular}{llccccccccc}
\toprule
\textbf{}    & \textbf{}     & \multicolumn{4}{c}{\textbf{Longitudinal Jerk} $k_x$}            &  & \multicolumn{4}{c}{\textbf{Lateral Jerk} $k_y$}                 \\ \cline{3-6} \cline{8-11} 
\textbf{Statistic} & \textbf{Avg} & \textbf{Mean} & \textbf{SD} & \textbf{Min} & \textbf{Max} &  & \textbf{Mean} & \textbf{SD} & \textbf{Min} & \textbf{Max} \\ 
\midrule
Absmax       & mean          & 19.611        & 1.997       & 14.356       & 24.020       &  & 33.401        & 3.525       & 25.763       & 46.110       \\
             & median        & 17.355        & 1.657       & 12.977       & 20.726       &  & 29.459        & 3.300       & 24.095       & 38.331       \\
Max          & mean          & 17.024        & 1.721       & 12.698       & 20.795       &  & 29.589        & 3.230       & 22.543       & 39.063       \\
             & median        & 15.483        & 1.452       & 11.599       & 19.833       &  & 26.604        & 2.905       & 20.587       & 33.588       \\
Min          & mean          & -17.215       & 1.859       & -21.722      & -12.794      &  & -29.236       & 3.108       & -38.831      & -23.955      \\
             & median        & -15.390       & 1.481       & -18.723      & -12.550      &  & -26.255       & 2.977       & -32.913      & -21.127      \\
SD           & mean          & 4.756         & 0.372       & 3.831        & 5.432        &  & 8.195         & 0.720       & 6.866        & 9.956        \\
             & median        & 4.652         & 0.352       & 3.822        & 5.577        &  & 7.887         & 0.688       & 6.762        & 9.536        \\
RMS          & mean          & 4.760         & 0.373       & 3.834        & 5.437        &  & 8.198         & 0.720       & 6.869        & 9.962        \\
             & median        & 4.654         & 0.352       & 3.822        & 5.580        &  & 7.889         & 0.689       & 6.762        & 9.544        \\
IDR          & mean          & 5.561         & 0.391       & 4.695        & 6.312        &  & 9.387         & 0.787       & 7.565        & 11.129       \\
             & median        & 5.441         & 0.376       & 4.570        & 6.422        &  & 9.014         & 0.797       & 7.251        & 11.416       \\
\bottomrule
\end{tabular}
\label{tab:jerkKpiResults}
\end{table}

\endgroup

\begingroup

\setlength{\tabcolsep}{3pt} 

\begin{table}[]
    \caption{Statistics of the distance to lane center and relative drift indicators.}
    \centering
    \scriptsize
\begin{tabular}{llccccccccc}
\toprule
\textbf{}    & \textbf{}     & \multicolumn{4}{c}{\textbf{Distance to Lane Center} $d_{\mathrm{CL}}$}                         & \textbf{} & \multicolumn{4}{c}{\textbf{Relative Drift Velocity} $v_{\mathrm{drift}}$}                       \\ \cline{3-6} \cline{8-11} 
\textbf{Statistic} & \textbf{Avg} & \textbf{Mean} & \textbf{SD} & \textbf{Min} & \textbf{Max} & \textbf{} & \textbf{Mean} & \textbf{SD} & \textbf{Min} & \textbf{Max} \\
\midrule
Absmax       & mean          & 0.398         & 0.058       & 0.277        & 0.552        &           & 1.605         & 0.175       & 1.198        & 1.975        \\
             & median        & 0.384         & 0.055       & 0.262        & 0.509        &           & 1.452         & 0.163       & 1.147        & 1.880        \\
Max          & mean          & 0.095         & 0.048       & 0.020        & 0.241        &           & 1.212         & 0.137       & 0.887        & 1.506        \\
             & median        & 0.086         & 0.054       & -0.003       & 0.222        &           & 1.070         & 0.137       & 0.747        & 1.349        \\
Min          & mean          & -0.350        & 0.063       & -0.513       & -0.192       &           & -1.436        & 0.172       & -1.852       & -1.100       \\
             & median        & -0.350        & 0.061       & -0.476       & -0.200       &           & -1.300        & 0.157       & -1.654       & -1.006       \\
Mean         & mean          & -0.149        & 0.053       & -0.272       & 0.014        &           & -0.104        & 0.062       & -0.261       & 0.032        \\
             & median        & -0.149        & 0.055       & -0.268       & 0.023        &           & -0.073        & 0.053       & -0.196       & 0.035        \\
SD           & mean          & 0.128         & 0.018       & 0.086        & 0.163        &           & 0.752         & 0.084       & 0.568        & 0.952        \\
             & median        & 0.123         & 0.018       & 0.092        & 0.161        &           & 0.657         & 0.070       & 0.522        & 0.851        \\
RMS          & mean          & 0.243         & 0.042       & 0.154        & 0.348        &           & 0.790         & 0.090       & 0.610        & 1.002        \\
             & median        & 0.226         & 0.040       & 0.142        & 0.314        &           & 0.689         & 0.074       & 0.529        & 0.881        \\
IDR          & mean          & 0.195         & 0.029       & 0.132        & 0.248        &           & 1.142         & 0.146       & 0.854        & 1.467        \\
             & median        & 0.181         & 0.030       & 0.113        & 0.236        &           & 0.927         & 0.109       & 0.775        & 1.192        \\
Median       & mean          & -0.154        & 0.056       & -0.290       & 0.027        &           & -0.096        & 0.070       & -0.271       & 0.019        \\
             & median        & -0.153        & 0.060       & -0.279       & 0.042        &           & -0.058        & 0.053       & -0.209       & 0.035        \\
\bottomrule
\end{tabular}
\label{tab:d2clVDriftKpiResults}
\end{table}

\endgroup

\begingroup

\setlength{\tabcolsep}{3pt} 

\begin{table}[]
    \caption{Class percentages and curve cutting intensities of the transient cornering behavior.}
    \centering
    \scriptsize
\begin{tabular}{lccclccc}
\toprule
\textbf{}             & \multicolumn{3}{c}{\textbf{Class Percentage}} & \textbf{} & \multicolumn{3}{c}{\textbf{Intensity} $\acute{I_T}$}     \\ \cline{2-4} \cline{6-8} 
\textbf{Trajectory Class}        & \textbf{Mean}  & \textbf{SD}  & \textbf{Max}  & \textbf{} & \textbf{Mean} & \textbf{SD} & \textbf{Max} \\ 
\midrule
Center                & 22.36          & 10.73        & 55.00         &           & 0.87          & 2.57        & 12.66        \\
Early   Cutting       & 4.17           & 4.58         & 20.00         &           & 17.33         & 27.68       & 166.45       \\
Early   Counter       & 4.81           & 4.71         & 17.39         &           & 26.83         & 80.42       & 612.72       \\
Late Cutting        & 6.90           & 6.27         & 24.00         &           & 29.07         & 37.03       & 178.55       \\
Late Counter        & 1.17           & 2.19         & 8.33          &           & 6.04          & 15.07       & 83.12        \\
Cutting               & 16.47          & 9.87         & 37.50         &           & 82.98         & 197.17      & 1576.52      \\
Counter               & 2.78           & 3.83         & 12.50         &           & 18.23         & 40.29       & 252.54       \\
Severe Cutting      & 1.86           & 3.67         & 16.67         &           & 22.28         & 57.70       & 281.86       \\
Severe Counter      & 0.28           & 1.55         & 10.00         &           & 1.15          & 6.55        & 44.86        \\
Biased Inner        & 31.96          & 13.99        & 66.67         &           & 217.03        & 114.58      & 625.94       \\
Biased Outer        & 1.46           & 2.90         & 13.04         &           & 55.59         & 122.60      & 450.57       \\
Oscillating           & 0.49           & 1.38         & 5.26          &           & 2.85          & 9.00        & 46.14        \\
Oscillating Cutting & 2.79           & 4.22         & 20.00         &           & 28.57         & 54.08       & 269.82       \\
Oscillating Counter & 1.03           & 2.41         & 11.11         &           & 16.08         & 55.46       & 297.45       \\
Slow Severe Cutting & 0.31           & 1.24         & 6.67          &           & 1.19          & 5.16        & 28.80        \\
Slow Severe Counter & 0.08           & 0.61         & 4.76          &           & 0.45          & 3.57        & 28.09        \\
\bottomrule
\multicolumn{8}{l}{\footnotesize{For percentage, all min values are zero except for the center class ($=5\%$).}} \\
\multicolumn{8}{l}{\footnotesize{Intensity values are multiplied by \num{1000} to account for low values.}} \\
\end{tabular}
\label{tab:TRKpiResults}
\end{table}

\endgroup

\begingroup

\setlength{\tabcolsep}{3pt} 

\begin{table}[h]
\caption{Loadings of the mean \ac{pca} of the proposed G-G envelope.}
 \centering
 \scriptsize
\begin{tabular}{lccl}
\toprule
\textbf{}      & \multicolumn{2}{c}{\textbf{Loadings}} & \textbf{}           \\ \cline{2-3}
\textbf{Angle} & \textbf{PC1}      & \textbf{PC2}      & \textbf{Uniqueness} \\ \midrule
-165           & 0.28              & 0.64              & 0.51                \\
-150           & 0.31              & 0.69              & 0.43                \\
-135           & 0.32              & 0.76              & 0.32                \\
-120           & 0.26              & 0.76              & 0.35                \\
-105           & 0.38              & 0.57              & 0.54                \\
-90            & 0.46              & 0.36              & 0.65                \\
-75            & 0.50              & 0.38              & 0.60                \\
-60            & 0.27              & 0.67              & 0.48                \\
-45            & 0.05              & 0.66              & 0.57                \\
-30            & -0.05             & 0.76              & 0.42                \\
-15            & 0.42              & 0.49              & 0.59                \\
0              & 0.48              & 0.61              & 0.40                \\
15             & 0.46              & 0.52              & 0.52                \\
30             & 0.73              & 0.37              & 0.32                \\
45             & 0.87              & 0.21              & 0.20                \\
60             & 0.91              & 0.21              & 0.13                \\
75             & 0.89              & 0.24              & 0.15                \\
90             & 0.92              & 0.16              & 0.13                \\
105            & 0.91              & 0.17              & 0.14                \\
120            & 0.86              & 0.23              & 0.21                \\
135            & 0.82              & 0.22              & 0.28                \\
150            & 0.77              & 0.42              & 0.24                \\
165            & 0.60              & 0.47              & 0.42                \\
180            & 0.62              & 0.40              & 0.46                \\ \bottomrule
\multicolumn{4}{l}{\footnotesize{Note: Varimax rotation was used}} \\
\end{tabular}
\label{tab:pcaMeanLoadings}
\end{table}

\endgroup

\end{document}